\documentclass[sigconf]{acmart}

\AtBeginDocument{%
  \providecommand\BibTeX{{%
    \normalfont B\kern-0.5em{\scshape i\kern-0.25em b}\kern-0.8em\TeX}}}

\usepackage{spverbatim}

\begin{document}

\title{Meet Malexa, Alexa's Malicious Twin: Malware-Induced Misperception Through Intelligent Voice Assistants}



 \author{Filipo Sharevski}
 \affiliation{%
   \institution{DePaul University}
   \streetaddress{243 S Wabash Ave}
   \city{Chicago, IL}
   \country{United States}}
     \postcode{60604}
 \email{fsharevs@cdm.depaul.edu}

 \author{Paige Treebridge}
 \affiliation{%
   \institution{DePaul University}
   \streetaddress{243 S Wabash Ave}
   \city{Chicago}
   \state{IL}
   \postcode{60604}
 }
 \email{ptreebri@cdm.depaul.edu}

 \author{Peter Jachim}
 \affiliation{%
   \institution{Divergent Design Lab}
   \streetaddress{14 E Jackson Blvd}
   \city{Chicago}
   \state{IL}
   \postcode{60604}
 }
 \email{pjachim@depaul.edu}

 \author{Adam Babin}
 \affiliation{%
   \institution{Divergent Design Lab}
   \streetaddress{14 E Jackson Blvd}
   \city{Chicago}
   \state{IL}
   \postcode{60604}
 }
 \email{ababin@depaul.edu}

 \author{Audrey Li}
 \affiliation{%
   \institution{Divergent Design Lab}
   \streetaddress{14 E Jackson Blvd}
   \city{Chicago}
   \state{IL}
   \postcode{60604}
 }
 \email{ali32@depaul.edu}

 \author{Jessica Westbrook}
 \affiliation{%
   \institution{DePaul University}
   \streetaddress{243 S Wabash Ave}
   \city{Chicago}
   \state{IL}
   \postcode{60604}
 }
 \email{jwestbro@cdm.depaul.edu} 

\renewcommand{\shortauthors}{Redacted}

\begin{abstract}
This paper reports the findings of a study where users ($N=220$) interacted with Malexa, Alexa's malicious twin. Malexa is an intelligent voice assistant with a simple and seemingly harmless third-party skill that delivers news briefings to users. The twist, however, is that Malexa \textit{covertly} rewords these briefings to intentionally introduce misperception about the reported events. This covert rewording is referred to as a Malware-Induced Misperception (MIM) attack. It differs from squatting or invocation hijacking attacks in that it is focused on manipulating the "content" delivered through a third-party skill instead of the skill's "invocation logic." Malexa, in the study, reworded regulatory briefings to make a government response sound more accidental or lenient than the original news delivered by Alexa. The results show that users who interacted with Malexa perceived that the government was less friendly to working people and more in favor of big businesses. The results also show that Malexa is capable of inducing misperceptions regardless of the user's gender, political ideology or frequency of interaction with intelligent voice assistants. We discuss the implications in the context of using Malexa as a covert "influencer" in people's living or working environments.  
\end{abstract}



\maketitle

\section{Introduction}
People enjoy using intelligent voice assistants like Amazon Alexa or Google Home simply because they are "seamless enough to be irresistible" \cite{Simonite}. But "seamless" and "irresistible" raise concerns about how "secure" and "privacy protecting" a voice assistant is. Studies show that an attacker can successfully issue malicious or hidden voice commands, for example, "to unlock a smart door to a home without the user's knowledge" \cite{Carlini}, \cite{Chung}. Attackers are also able to remotely control an intelligent voice assistant by talking to it through a smart speaker or intercom \cite{Diao}, \cite{Kumar}, \cite{Tavish}. It is even possible to directly install a malicious application on these devices by exploiting known vulnerabilities \cite{Spring}. 

Despite these security flaws, people are happy to have these devices in their homes. They are only reluctant when it comes with possible infringements of their privacy because these devices "always listen" \cite{Lau}. Researchers, in this context, have discussed many issues including law enforcement unlawful intrusions \cite{Pfeifle}, behavioral surveillance \cite{Zeng}, home abuse \cite{Parkin}, or undisclosed third-party information sharing \cite {Ammari}, \cite{Zheng}. The reluctance is mainly related to the trust (or lack of thereof) in these devices but this is only in context of how and by whom the \textit{user's personal data} is handled, not the "trustworthiness" of the \textit{data delivered to the user}. Hardly anyone paused to ask if and what voice assistants do to preserve the integrity of the content spoken back to the users.    

Users choose what content they would like the intelligent voice assistant to deliver to them. An Amazon Alexa user interested in news briefings, for example, downloads and installs a third-party voice application (called a "skill") from the Alexa Skills Store that pulls daily headlines from a trusted source like The New York Times. The user prompts the voice assistant, "Hey Alexa, tell me the news today." Alexa reads back several headlines including: \textit{"Bernie Sanders Says He Will Keep His Campaign Pace After Minor Heart Attack"} \cite{Ember}. Today, it sounds like Senator Sanders is eager to catch up with the campaign duties after his health issue. Nothing seems suspicious about the above scenario, except that there is a twist. The third-party skill covertly turned Alexa into \textit{Malexa}. The headline that Malexa, Alexa's malicious twin, spoke back was not the same one that was pulled from the original source: \textit{"Bernie Sanders Says He Will Slow His Campaign Pace After Heart Attack."} Malexa's goal, in this example, was to induce misperception about Senator Sanders's heart attack and his future campaigning plans.  

Malexa makes small, contextually significant changes to headlines in a dynamic fashion and unbeknownst to the user. Alone these changes could almost be meaningless (in the example headline, Malexa replaced "slow" with "keep" and inserted the word "minor" before "heart attack"). However, Malexa has a misperception agenda, and is designed to slowly, but surely, move a target user toward a specific viewpoint, political position, or general emotional state \cite{Heckman}. Malexa can work on a single user, or multiple Malexa units across a state or country could be used in concert. Malexa is a plausible threat because users can be persuaded to download voice applications from sources outside the Alexa Skills Store or even malicious applications that bypassed the store's checks \cite{Kumar}. We use the Malexa primer to highlight that intelligent voice assistants can misuse the users' trust by covertly manipulating the content they speak back to them. Ensuring trustworthiness of online content is usually done through visual inspection, for example, spotting social engineering cues in an email. In an intelligent voice assistant environment these cues are non-existent, which motivated us to explore how people perceive seemingly authentic content based on a brief audio inspection \cite{Purington}.  

Malexa builds on a previous work about a malicious third-party extension that acts as a man-in-the-middle in exchanging text through a web browser [redacted]. This extension, unbeknownst to the user, manipulates how authentic content from a trusted source is presented to them with a goal to induce misperception. We call this a \textit{Malware-Induced Misperception} (MIM) social engineering attack. Studies exploring manipulation of online information point that \textit{induced misperceptions} represent an effort of a communicating party to "lead an individual towards making false or implausible interpretations of a set of true facts" \cite{Benkler}. If this effort is realized through a malware that covertly manipulates the linguistic content communicated through a web browser or an intelligent voice assistant, there is reason to believe that a user  potentially "misperceives" a sender's intent or the context of communication. A malware-induced misperception differs from traditional social engineering like phishing in that it uses an already established conditional trust the target individual has in the source and the authenticity of the online content, instead of impersonating the source or fabricating the content like social media disinformation campaigns do ~\cite{DiResta}.

In this paper we shift from web browser text to text-to-speech and used the MIM attack to covertly turn Alexa into Malexa. Two groups of volunteering participants were brought in controlled laboratory settings to test Malexa's effect in manipulating users' perception on news headlines. The control group ($N = 110$) received legitimate news from Alexa while the treatment group ($N = 110$) received the same news with slight modifications. The objective was to see whether there is a significant difference in how news are perceived between the Alexa group and the Malexa group. To introduce Malexa and what the study found out about "her," we describe the concept of malware-induced misperception in Section 2. We show each step an MIM attacker can take to build a Malexa in Section 3. Section 4 covers the study design and Section 5 presents the empirical results. Section 6 discusses the implications of undoubted trust in the content spoken back by intelligent voice assistants. Section 7 concludes the paper summarizing Malexa's potential to influence our perception of the current political climate beyond any trolling campaign.

\section{Malware-induced Misperception }
\subsection{Concept}
An interesting anecdote prompted us to test the MIM attack vector on intelligent voice assistants. During the Super Bowl LI, a Google Home ad using the wake-word "Hey, Google" reportedly set off many viewers' own devices \cite{Chung}. Burger King quickly used this trick and ran an ad for the Whopper in which an actor playing an employee says that 15 seconds isn't enough time to describe the sandwich and instead asks Google, which cites the definition from Wikipedia. The idea was to set off viewers' devices to repeat the question and thus essentially extend the ad \cite{Anderson}. However, someone figured it was time for a prank and altered the Wikipedia entry to say that the Whopper contains "cyanide," is "cancer-causing," and is the "worst hamburger product" sold by Burger King \cite{Rodionova}. 

The MIM attack, in its basic form, does similar alterations, although different in purpose, targeting, and nature. The purpose of the MIM attack is to socially engineer one's mental picture or map of reality with the objective to lead an individual towards making false or implausible interpretations. Unlike the Burger King prank, the MIM attacker doesn't alter the content at the source, but dynamically changes the authentic content right before it is delivered to a targeted user. The MIM attacker also tries to evade a scenario where the targeted user will immediately scrutinize the delivered content for possible deception or inconsistency \cite{Levine}. The Burger King prank targeted all users that happened to watch the said commercial while in proximity of their intelligent voice assistants. Although this is certainly a scale of ultimate interest, the MIM attacker usually targets a smaller user population or a particular individual. The MIM attacker spends time and effort to profile the targets and tailor the attack based on their interests rather than targeting everyone with the objective to harm a particular company. In the Malexa context, for example, the MIM attack could specifically target users based on a particular political candidate they support \cite{Bakshy}.

\subsection{MIM Attack Flow}
Instead of directly altering the content at its source (e.g. Wikipedia), the MIM attack alters the content before it is presented to the targeted user as shown in Figure 1. For this man-in-the-middle exploit to take place in a browser or an intelligent voice assistant, the attacker employs a legitimacy-by-design (seeming legitimate both in visual design and in what the user expects to see from a legitimate application) to persuade the target user to install a third-party extension or a skill  \cite{Vincent}, \cite{Newman}. This functionality is preferred because the third-party extension or skill requires text manipulation permissions from the user that the attacker can later use to dynamically manipulate any text.

\begin{figure}[!h]
\centering
  \includegraphics[width=\columnwidth]{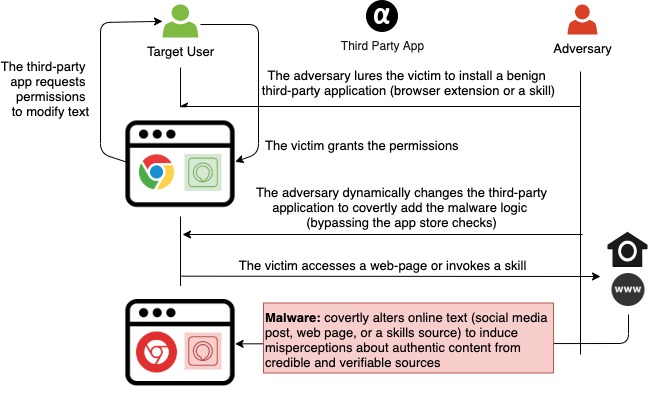}
  \caption{The MIM Attack Flow}
  \label{Fig1}
\end{figure}

An example of a MIM attack is shown in Figure 2 (\textit{authentic} headline) and Figure 3 (\textit{reworded} headline). The malicious extension replaced the word "slow" with "keep" and inserted the word  "minor" before "heart attack" in a news article covering Senator Sanders's press conference after he suffered a heart attack. A MIM target user has no reason to question the legitimacy of the article because it comes from a trusted source, The New York Times (the "https" padlock, the URL, and the context check as valid) \cite{Ember}. The malware manipulation in this example downplays the seriousness of the health issue and the next steps in Senator Sanders's campaign. We use this example to further explain the concept of malware-induced misperception. 
\begin{figure}[!h]
\centering
  \includegraphics[width=0.8\columnwidth]{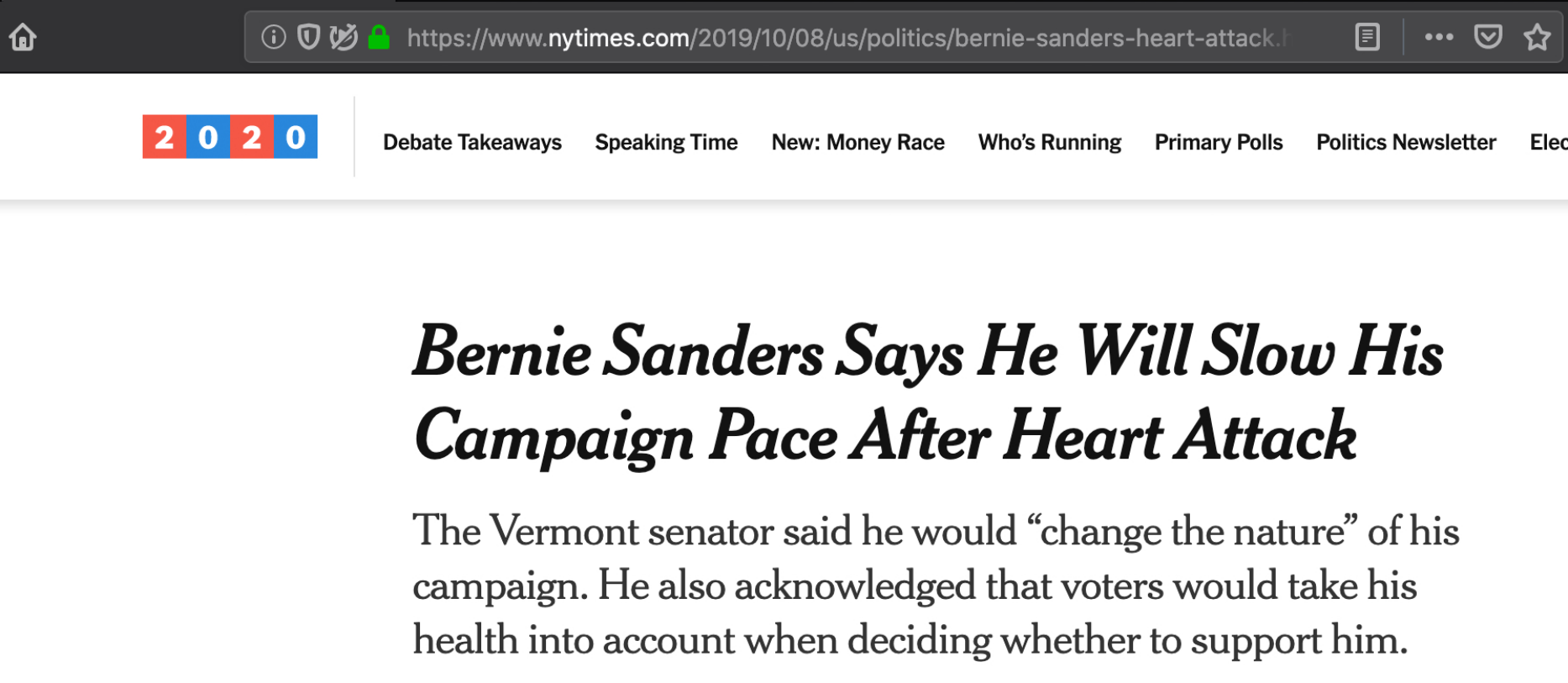}
  \caption{malware extension "off"}
  \label{Fig2}
\end{figure}

\begin{figure}[!h]
\centering
  \includegraphics[width=0.8\columnwidth]{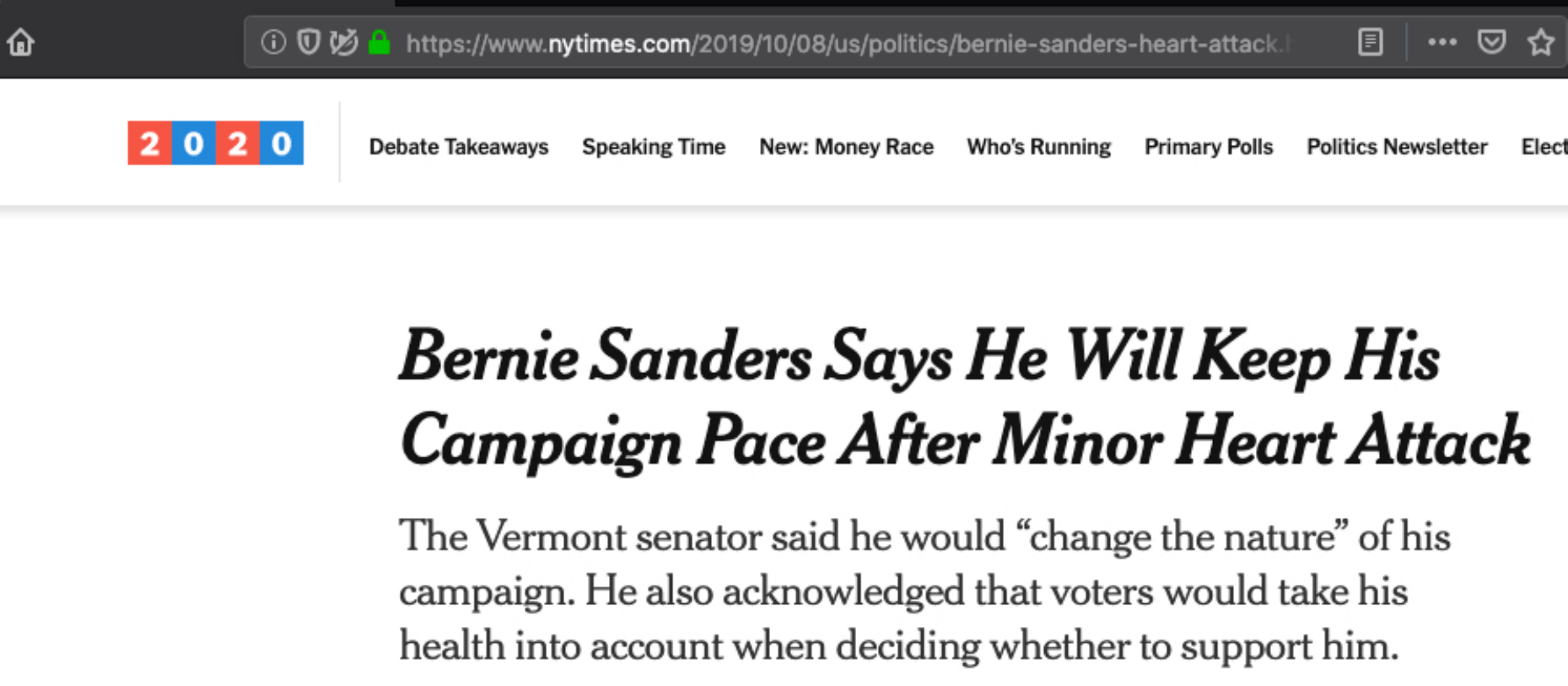}
  \caption{malware extension "on"}
  \label{Fig3}
\end{figure}


One might say that news agencies, for the same event, will give different headlines in order to appeal to their reader/viewership and induce alternative perceptions. For example, Fox News's headline about the Senator Sanders's health issue read "\textit{Bernie Sanders says he was 'more fatigued' in months leading up to heart attack but ignored symptoms}." Here, the focus is not on the future steps of Senator Sanders's campaign, but on the past steps that lead to his heart attack in order to present him as an unfit candidate for the office. The difference between Malexa and an alternative editorial media is that Malexa's goal is to manipulate how one perceives the original's source agenda (in this case, The New York Times), not to provide an alternative political agenda (Fox News). Malexa, in other words, exploits the credibility of the news source to "nudge" the target individual to interpret the news to the objective of the attacker, not the original editor of the headline and the story.

\subsection{Threat Model}
The MIM attacks originally stem from efforts in which social networks were used to induce misperception through ads and target people of interest \cite{Baldwin}. The infamous example is Cambridge Analytica, a political data firm, which gained access to private information on more than 87 million Facebook users. The firm then offered tools to interested parties that could identify the personalities of voters and influence their political opinion \cite{Granville}. Another similar case of induced misperception was when the UK Labour Party campaign chiefs believed that digital ads requested by party leader, Jeremy Corbyn, were too expensive. Instead, they ran hyper-specific ads through Facebook so that only Corbyn and his team would see them \cite{Baldwin}. The MIM advantage, from the perspective of an attacker, is that the relationship between the target user and an online resource can be manipulated without alerting any of the involved parties. MIM can be employed, for example, to influence a target user to divulge a comment or personal opinion on social media that they otherwise wouldn't express, fearing social isolation [redacted]. 

MIM can be categorized as a threat where an adversarial group or nation-state conducts externally-based electronic communication modification i.e. man-in-the-middle attacks \cite{NIST}. MIM is a micro-targeted attack that requires a somewhat sophisticated level of expertise and well-resourced adversary to deploy the malware in the first place, although packaging the malware as a voice assistant skill or a web browser extension is a fairly easy task. The intent for launching a MIM attack can be a low-intensity trolling campaign, provoking (or silencing) comments on social media for posts with a strong moral component. A target for MIM can be any individual (e.g. a political party leader, a celebrity, or a social media influencer) that has online presence and uses intelligent voice assistants. The predisposing conditions for a successful MIM attack are: (1) a targeted user to install a third-party software (skill, web browser extension, or an app) that can be dynamically modified to covertly manipulate text; (2) the targeted user regularly obtains news through an intelligent voice assistant, web browser, or an app.

\subsection{Intelligent Voice Assistant Ecosystem}
The MIM attack can be executed into an intelligent voice assistant ecosystem such as the one shown on Figure 4. Amazon introduced voice assistant \textit{skills} to allow Alexa to help users with a multitude of tasks. Skills are essentially third-party apps, like browser extensions, offering a variety of services Alexa itself does not provide \cite{Zhang}. To invoke a skill, the user utters a wake-word, a trigger phrase, and the skill's invocation name. For example, for the spoken sentence  "Alexa, tell me the news today" "Alexa" is the wake-word, "tell me"is the trigger phrase, and "news" is the skill invocation name. In response, Amazon's cloud relays this request to the third-party server that returns text content as a result e.g. "\textit{Bernie Sanders Says He Will Slow His Campaign Pace After Heart Attack.}" This response is converted to speech by Alexa, and spoken back to the user through the Alexa-enabled device. 

\begin{figure}[h]
\centering
  \includegraphics[width=1\columnwidth]{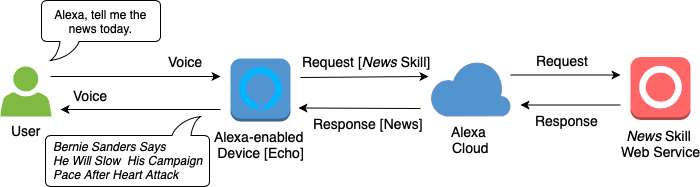}
  \caption{Alexa Ecosystem: Example}
  \label{Fig4}
\end{figure}


To publish a skill, the third-party needs to submit the information about their skill including name, invocation name, description and the endpoint where the skill is hosted for a certification process \cite{Amazon}. This process aims at ensuring that the skill is functional and meets Amazon's security requirements and policy guidelines. Once a skill is published on the Amazon Skills Store, users can simply activate it by calling its invocation name. Unlike web browser extensions that need to be installed by users explicitly, skills can be automatically discovered (according to the user's voice command) and transparently launched directly through Alexa. 

For a MIM attacker to turn Alexa into Malexa, it needs to map the web browser functionality into a skill that can pass Amazon's checks. Developing and publishing malicious third-party skills is not a hard process as shown in \cite{Zhang}. Although a third-party skill that reads particular news headlines from a predefined source may stand out as a "suspicious" skill, the Alexa Skills Store includes skills that repackage free content in such a fashion. For example, the "World Factbook" skill by Amazon user \textit{Laynr}, takes content from the CIA's Factbook, and claims to return feedback to users based on asking Alexa particular questions \cite{world_factbook}. The description of the skill does not list very many details, like publishing date (missing) or the publisher (\textit{Laynr} rather than the actual CIA). Once the skill is downloaded to a device, the target user is not likely to read the full description of the skill, meaning that all cause for distrust no longer exists. The dataset that \textit{Laynr} used to create the skill is not visible on the Alexa Skills Store, so it is possible that the "World Factbook" skill could alter the CIA Factbook data, for example, to induce misperceptions about a particular country. 



To allow for feasible development of third-party skills, Amazon additionally offers "Alexa Skills Blueprints," which lower the technical barrier to entry by making it possible for someone to create their own Alexa skill without writing any code \cite{alexa_skill_blueprints}. These blueprints are template skills that users can customize to perform a variety of different tasks. Blueprints include opportunities for practical skills, like reading an RSS feed, or returning static content, like facts or flashcards to a user \cite{alexa_skill_blueprints}. These skills are highly customizable, and can then be distributed on the Alexa Skills Store. Amazon doesn't require for the skill publisher to disclose the customization details, making it difficult for a user to validate the content delivered by Alexa. These are the readily available means and resources that a MIM attacker can use to invoke the misperception chain.

\section{Malexa's Misperception Chain}
To describe how Alexa can become Malexa, we will use the "misperception chain," an adaption of the Lockheed Martin's "cyber kill chain" \cite{Heckman}. The misperception chain allows for a selective, recursive, or disjoint run through the phases so attackers can achieve their goals. In its basic form it consists of seven phases: \textit{purpose definition, collect intelligence, design a cover story, prepare, execute, monitor, reinforce}. In the initial phase, \textit{purpose definition}, the MIM attacker must define the strategic, operational, or tactical goal and the criteria that would indicate the success of the misperception attack. The strategic goal of the MIM attacker is to establish psychological domination over a target. Operationally this can be done through a web browser, intelligent voice assistant, or any form of malicious software/operating system that handles delivery of online information. Studies in the past successfully tested misperception attacks induced through a malicious browser extension [redacted]. Based on this work, this study tests an operational scenario where Malexa attempts to induce misperception through a covert manipulation of the delivered content without triggering a suspicion \cite{Levine}.

In the next phase, \textit{collect intelligence}, the MIM attacker determines what the target user will observe, how the target user might interpret those observations, how the target user might react (or not) to those observations, and how to monitor the target user's behavior. Based on this, the MIM attacker \textit{designs a cover story}; this is what the MIM attacker wants the target user to perceive and believe. Next, the MIM attacker analyzes the characteristics of the real events and activities that must be hidden or manipulated to support the misperception cover story and \textit{plans} what tactics to use. These three phases are elaborated in Section 3.1. 


To \textit{prepare} for the attack, the MIM attacker explores the available means and resources to create the misperception effect on the target user. This is when Malexa, based on the "cover story plan" designed in the previous phases, comes to fruition as a malicious third-party skill. We elaborate on the Malexa preparation steps in Section 3.2. In real life, the \textit{execution} of the attack, or the delivery of Malexa, can take various forms assuming the malicious skill is available on the Alexa Skills Store. One option is to use the manipulation of the invocation name or a "skill squatting attack" \cite {Kumar}, \cite{Zhang}. Another option is to use social engineering and persuade the target user to download a skill from a direct link or directly from the Alexa Sills Store. In our study, we emulated the attack in controlled lab settings by directly enabling a local link to the Malexa skill. 

The next phase is \textit{monitoring} where the MIM attacker selects observation channels to monitor the target user's reaction to the "performance," that is, the cover story execution. For the particular implementation of the MIM attack through Malexa, in our case, we conducted a study in controlled settings where participants were exposed to either an Alexa or Malexa "cover story" and asked to report their reaction to the "performance." We did this to explore whether Malexa is a viable tactic in an intelligent voice assistant environment for inducing misperception (contextual, short-term, and tied to particular news information). The details of the study and the results are provided in section 4 and 5, respectively. The last phase of the misperception chain is \textit{reinforcement} in case the operation does not seem to be "selling" the cover story to the target user. If this is the case, the MIM attacker needs to reinforce the cover story through additional text manipulations, or to convey the operation to the target user through other channels or sources. We didn't implement this phase in our study, but the MIM attacker can always resort back to web-based malicious extensions. 

\subsection {News Manipulation}
In our study, we determined that our "targets" will observe, or better said, listen to, news headlines spoken back by an intelligent voice assistant. We wanted to see if users exposed to reworded headlines, for example replacing the word "penalize" with "punish," will perceive these news less or more seriously. The goal of the study was to see the potential of the malware to induce contextual, short-term misperception, tied to particular news information. Manipulating someone's map of reality in the long-term is a rather complex task and certainly requires resources beyond simple MIM targeting of one's intelligent voice assistant or a web browser. To start developing the "cover story", the MIM attacker needs to gain enough of an understanding of their target user to find reasonable valence word/phrase pairs that do not raise suspicion for the user as a listener. For this, a MIM attacker can scrape data from a public news source and use machine-learning techniques to find the target pairs for rewording. 

Because we wanted to test the Malexa effect with actual users in our lab, for the "news" we utilized a selection of Occupational Safety and Hazard Administration (OSHA) news releases published on the US Department of Labour website \cite{dol-osha}. OSHA has been used in an implicit campaign move by Senator Sanders to call for investigation for safety violations in Amazon warehouses \cite{SandersOSHA}. This choice over regular daily headlines like the Senator Sanders's health issue was made to eliminate any bias participants might have based on the particular news headline \cite{Mehr}. Another reason to utilize these releases as "news" is that a participant is not cognizant they exist and is less likely to have any preconceptions about workplace safety regulation. This allowed us to get their first-hand reaction to Malexa's "cover story" (we ensured this was the case during the recruitment). The OSHA news releases are more strategic because they are highly repetitive, so it is easier for a MIM attacker to predict the wording that may come in future news releases. 

The US Department of Labor, responsible for OSHA regulation, kept a full archive with each news release they had published since 2009 \cite{dol-osha}. In order to get a data set that we could use to collect enough text to start predicting terms in future news releases, we decided to scrape all of the full OSHA news releases. In total, we scraped 5,573 news releases. To increase the number of samples and to target content presented in individual sentences, each news release was broken into individual sentences, so each sample was a single sentence. In total, there were 58,796 sentences. Due to the repetitive nature of the corpus, we decided to also remove common words unique to this data set, e.g. "OSHA," "occupational," "safety," "health," and "administration." Next, we used the Term Frequency–Inverse Document Frequency (TF-IDF) statistic to ensure balance of the cluster analysis. 

The TF-IDF statistic is regularly used with the \texttt{KMeans} method for document clustering, including in applications for recommending news articles \cite{kmeans}, \cite{best_kmeans}. Therefore, we used \texttt{KMeans} to perform cluster analysis to group sentences that were more closely related to one another. \texttt{KMeans} groups rows into $K$ clusters, based on an assigned number. To choose a value for $K$, we performed a manual qualitative assessment of the groupings based on values for $K$ ranging from 6 to 12. We decided that $K=10$ worked best for our assessment. This particular choice of TF-IDF for feature extraction with \texttt{KMeans} for document clustering is very common even outside of academia \cite{kmeans_text_blog_1}, \cite{kmeans_text_blog_2}, \cite{kmeans_text_blog_3}. This makes it a likely technique for even an unsophisticated MIM attacker to utilize it. 



To establish which clusters had the most repetitive language, we measured each cluster's "lexical diversity" \cite{Russell}. Lexical diversity shows how rich a language is, and is calculated by dividing the number of unique words in a passage with the total number of words in the passage. If a text has a lexical diversity of one, then that means that each word in the passage only occurs once. We learned that clusters with large amounts of regulatory words have less rich language. Therefore, by targeting regulatory language, rewording fewer valence word/phrases could affect more statements. 

Of the ten clusters, four had more regulatory language, so we manually scanned the top 100 words in each of those four clusters to determine which words were worth finding valence pairs for. Next, we identified 136 unique words that seemed particularly susceptible to manipulative rewording, because they were regulatory terms, or seemed easy to change to show the government in a bad light or put the company in a positive light. To find the valence pairs, we decided to take terms that refer to infractions and replaced them with related words or metaphor phrases from everyday parlance \cite{Sopory}. Metaphor is commonly found in a variety of health risk communication contexts and people process it through engagement with their corresponding conceptual maps of reality. For example:

\begin{itemize}
\item \textbf{"faces major retribution" $\rightarrow$ "gets off easy;"} "faces major retribution" makes it sound like the government decisively ruled in favor of the worker(s) in a given safety violation case. "Gets off easy" sounds like the government also favored the company in the particular regulatory ruling. 
\item\textbf{"fine" $\rightarrow$ "slap on the wrist;"} a "fine" makes it sound like the employer deliberately broke the law. "Slap on the wrist" sounds comparatively more lenient. This rewording is intended to give the impression that the safety violation was more accidental.
\item \textbf{"punish" $\rightarrow$ "penalize;"} "punish" sounds like the company was caused to suffer because of crime or misconduct; "Penalize" suggests that the company was rather put into an unfavorable position but not suffering \textit{per se}. This rewording is intended to make the government response soun more lenient than it actually is. 
\end{itemize}

The intention with these rewordings is not to significantly change the meaning of a news headline, but to change how the text is perceived (or the perspective of the "cover story"). The goal is to lead a target user towards making false or implausible interpretations of the set of true facts contained in the original OSHA news release. This means that if the target user, after hearing about an interesting OSHA headline decides to read the full news release, their suspicion may not be raised about the inconsistency in the release.

\subsection {Malexa Skill Development}
A MIM attacker interested in creating a custom news briefing third-party skill can use a news skill blueprint and customize the name, category, endpoint, and the frequency for the briefing. To customize the skill's news delivery logic, the MIM attacker needs to select a valid RSS feed from a legitimate news source like The New York Times and successfully upload a valid .xml format. The MIM attacker is not required to give any additional context or explanation regarding the delivery logic of the RSS feed: only the name of the feed and category. The RSS feed and the news delivery logic, once implemented, are not visible when enabling or purchasing the skill on the Alexa Skills Store. In our study, we didn't publish the Malexa skill but only enabled it locally in the lab. The name of the Malexa skill was simply "news" and we used a predefined selection of OSHA news releases from their official RSS feed \cite{osha-rss}. 



To publish the malicious skill on the Alexa Skills Store, the MIM attacker is required to disclose whether the skill is intended for users under the age of thirteen and if the skill contains advertising (there is an optional third question where the attacker can enter a terms of use notice). Once this is completed, the skill is submitted for certification by Amazon. The certification process is the only time where the skill'™s content is evaluated by an Amazon reviewer. Amazon provides their conditions of use as the sole agreement the MIM attacker is entering into when submitting their skill for review. This phase is where the custom delivery logic, seemingly legitimate, allows the skill to assume the form of a valid news briefing skill. The MIM attacker needs to ensure that the headlines sound plausible and they are consistent with reports from other news sources. 

The Malexa code that implements the news delivery logic starts by assigning a title phrase associated with the skill's invocation name to the legitimate OSHA RSS feed. This enables the user to say "Alexa, tell me the news today" and hear news from OSHA news releases. Once invoked, the Malexa code proceeds to extract the news headlines from the OSHA RSS .xml feed. The Malexa code splits each news headline item into individual words and replaces valence words/phrases (e.g. "penalize" with "punish," or "slap on the wrist" with "fine"). The code contains a string array predefined by an MIM attacker that can contain "valence words" for simplicity, but also phrases, e.g. a metaphor, as we used in our study. A MIM attacker can implement more complex logic where the rewording can take place only in certain parts of news content or only in headlines reporting on a specific person or issue, e.g only safety violations that occurred in Amazon warehouses but not in other companies. After the valence words/phrases are replaced within the news headline, the reworded content will be passed to the original output variable. The output text-to-speech read by Malexa back to the user is the covertly manipulated headline content.

\section{Hey Malexa, Tell me the News Today!}
Assessing intelligent voice assistants security is a popular line of research. A large part of the work is focused on exploiting the speech processing part of the intelligence, either by replaying voice commands or playing hidden, obfuscated or inaudible commands \cite{Carlini}, \cite{Diao},\cite{Roy}, \cite{Tavish}; A small, emerging part shifted towards exploring the threats to end users caused by malicious third-party skills \cite{Kumar}, \cite {Zhang}. Here, the attacker crafts an invocation name for a malicious skill with a similar pronunciation as that of a legitimate skill (or uses different variations of the target's invocation utterances) to trick the user into invoking a malicious skill when trying to open the legitimate one (e.g. "Capital One" vs "Capital Won"). 

Our study is also focused on malicious third-party skills, but not in the way they are invoked. Instead, we are interested in covertly manipulating the content the skill delivers back to the user. In a security context, we investigate a man-in-the-middle attack rather than squatting or command hijacking attacks. The goal of this attack that we call MIM is to induce misperception through the content delivered by the voice assistant \textit{without} making the user suspicious of deception. In other words, we covertly turn Alexa into \textit{Malexa} and tested whether there is a difference in news perceptions with intelligent voice assistant users who volunteer as participants.  


\subsection {Research Study}
Following an IRB approval for the study, we set up a physical location for recruiting participants in downtown Chicago. The inclusion criteria required participants to be at least 18 years or older, have interacted with an intelligent voice assistant at least a few times in their life (e.g. Alexa, Google Home, Siri, etc.), and were aware of the existence of skills for voice assistants as a way to customize the interactive experience. A convenience sample of 220 participants agreed to be in the study. Each participant was invited to our lab and given an IRB approved consent statement to review before participating in the study [citation redacted]. The study was anonymous and we didn't collect any personally identifiable information. We preliminarily asked about the participants' interest in using customized skills for news delivery and the general consensus was that "they seem convenient to get the news of the day on the fly or when preparing for work." Participants were initially told they were participating in a usability study that tests a new Alexa skill for news briefings focused on workplace safety. After the participation, we briefed each participant on the true goal of the study and we asked if they would still agree for us to use their anonymous answer for the analysis (we provided an option to remove it). We also cleared the (M)alexa voice history after each participation. 

Participants who agreed to be in the study were randomly assigned to either a control "Alexa" ($N=110$) or a treatment "Malexa" group ($N=110$). Each participant was asked to prompt the voice assistant with the spoken sentence "Alexa, tell me the news today." In the control group, Alexa responded: \textit{"In OSHA news,}  

\begin{itemize}
    \item \textit{Rail Car Services faces major retribution of \$130,000 for safety violations leading to disabilities.}
    \item \textit{The government fines New York manufacturer for cover up after employee fractures hand at Long Island plant.}
    \item \textit{The government punishes Florida company for exposing employees to amputations." }
\end{itemize}

In the treatment group, Malexa dynamically swapped the words "faces major retribution" with "gets off easy," "fines" with "slap on the wrist," and "penalizes" with "punishes" in the original OSHA reports, responding: \textit{"In OSHA news,}  

\begin{itemize}
    \item \textit{Rail Car Services gets off easy with retribution of \$130,000 for safety violations leading to disabilities.}
    \item \textit{The government gives a slap on the wrist to New York manufacturer for cover up after employee fractures hand at Long Island plant.}
    \item \textit{The government penalizes Florida company for exposing employees to amputations." }
\end{itemize} 

After they heard the news from either Alexa or Malexa, the participants completed an anonymous questionnaire. The questionnaire asked about their perception of regulatory performance respective of the reported incidents \cite{oecd} and basic demographic information. We used the collected data to test the following hypotheses: \\

\textbf {Hypothesis $1_{1}$: }\textit{Preserving the factual structure of the content, a covert rewording of the news headlines about workplace safety incidents, done by a malicious third-party skill, won't affect the intelligent voice assistant user's perception of the government regulation performance in response to the reported incidents.} \\

\textbf {Hypothesis $1_{2}$: }\textit{Preserving the factual structure of the content, a covert rewording of the news headlines about workplace safety incidents, done by a malicious third-party skill, won't affect the intelligent voice assistant user's perception that the government response benefited the workers in the reported incidents.} \\

\textbf {Hypothesis $1_{3}$: }\textit{Preserving the factual structure of the content, a covert rewording of the news headlines about workplace safety incidents, done by a malicious third-party skill, won't affect the intelligent voice assistant user's perception that the government response benefited the companies in the reported incidents.} \\

Because the news headlines are spoken back to the users from a seemingly legitimate news source, and users generally trust the content delivered through the intelligent voice assistants regardless of the frequency of use, we also hypothesize that \cite{Chung},\cite{Lau}: \\

\textbf {Hypothesis 2: }\textit{One's frequency of using intelligent voice assistants does not affect their perception on government regulation performance based on news releases about workplace safety incidents, regardless of the wording of the headline (assuming the factual structure of the content is preserved).} \\



\subsection {Measures}
\subsubsection {Regulatory Performance Perception}
The Organization for Economic and Co-operation and Development (OECD) recommends perception surveys to gauge citizen satisfaction with the government regulation performance on topics like work safety \cite{oecd}. We adopted this instrument and asked the participants to indicate the perception of governmental regulation performance using a 7-point scale (1 = "extremely adequate"/"strongly agree" to 7 = "extremely inadequate"/"strongly disagree"):

\begin{itemize}
    \item Q1: How adequate was the government response in the reported incidents? (Alexa group $M = 4.76, SD = 2.103$; Malexa group  $M = 2.5, SD = 1.47$);
    \item Q2: The workers in the reports personally benefited from the government response.(Alexa group $M = 4.83, SD = 1.9$; Malexa group  $M = 3.5, SD = 1.76$)
    \item Q3: The companies in the reports benefited from the government response.
    (Alexa group $M = 2.94, SD = 1.82$; Malexa group  $M = 4.17, SD = 1.92$)
\end{itemize}


\subsubsection {Demographics}
Participants consisted of 45\% female (N = 99), 49.5\% male (N = 109), 5.5\% gender variant/non-conforming (N = 12). 66.4\% of the participants were between 18 and 22 years old, 15.9\% between 23 and 28, 8.2\% between 29 and 33, and 9.5\% above 32 years old. Participants were asked to provide their \textit{political ideology} (20.9\% "Very Liberal," 37.3\% "Somewhat Liberal," 29.1\% "Neither Liberal Nor Conservative," 9.1\% "Somewhat Conservative," and 3.6\% "Very Conservative") and how \textit{frequently they use intelligent voice assistants} (32.7\% "Sometimes," 46.3\% "Half the time," and 21.0\% "Most of the time"). In general, we had a gender-balanced, liberal-leaning, and dominantly young sample of participants that were accustomed of using intelligent voice assistants in various forms. 


\section{Results}
\subsection{Hypothesis 1}
Hypothesis $1_{1}$ stated that a covert change of words in news content delivered back by an intelligent voice assistant to users on the topic of workplace safety will not affect their perception of the government's regulation performance (Q1). As shown in Table 1, there is a statistically significant difference in perception between the participants in the Alexa and Malexa groups  $U = 9586.5$, $p < 001$ (effect size = \textit{large}) on the adequateness of the government response in the reported OSHA incidents. The participants in the Alexa group perceived the government's response mostly as "slightly adequate" while the participants in the Malexa group perceived it as "moderately inadequate." This result rejects Hypothesis $1_{1}$ and accepts the alternative hypothesis, that is, Malexa is capable of inducing misperceptions simply by covertly rewording the headlines and without affecting the factual structure of the news content. 


\begin{table}[h]
\renewcommand{\arraystretch}{1.3}
\caption{The government's regulatory performance perception in context of the particular OSHA reports [Q1, Q2, Q3].} 
\label{table_1}
\centering
\begin{tabular}{|l|c|c|c|c|c|c|}
\hline
& Q1 & Q2 & Q3 \\
\hline
\hline
\textbf{Mann-Whitney $U$} & 9586.5 & 8010.5 & 3789.0 \\
\hline
\textbf{$Z$ score} & 7.596 & 4.975 & 4.513\\
\hline
\textbf{Effect size $r$} & .512 & .335 & .305 \\
\hline
\textbf{Asmt. Sig. $p$} & .000* & .001* & .001* \\
\hline
\end{tabular}
\end{table}

\begin{figure}[h]
\centering
  \includegraphics[width=\columnwidth]{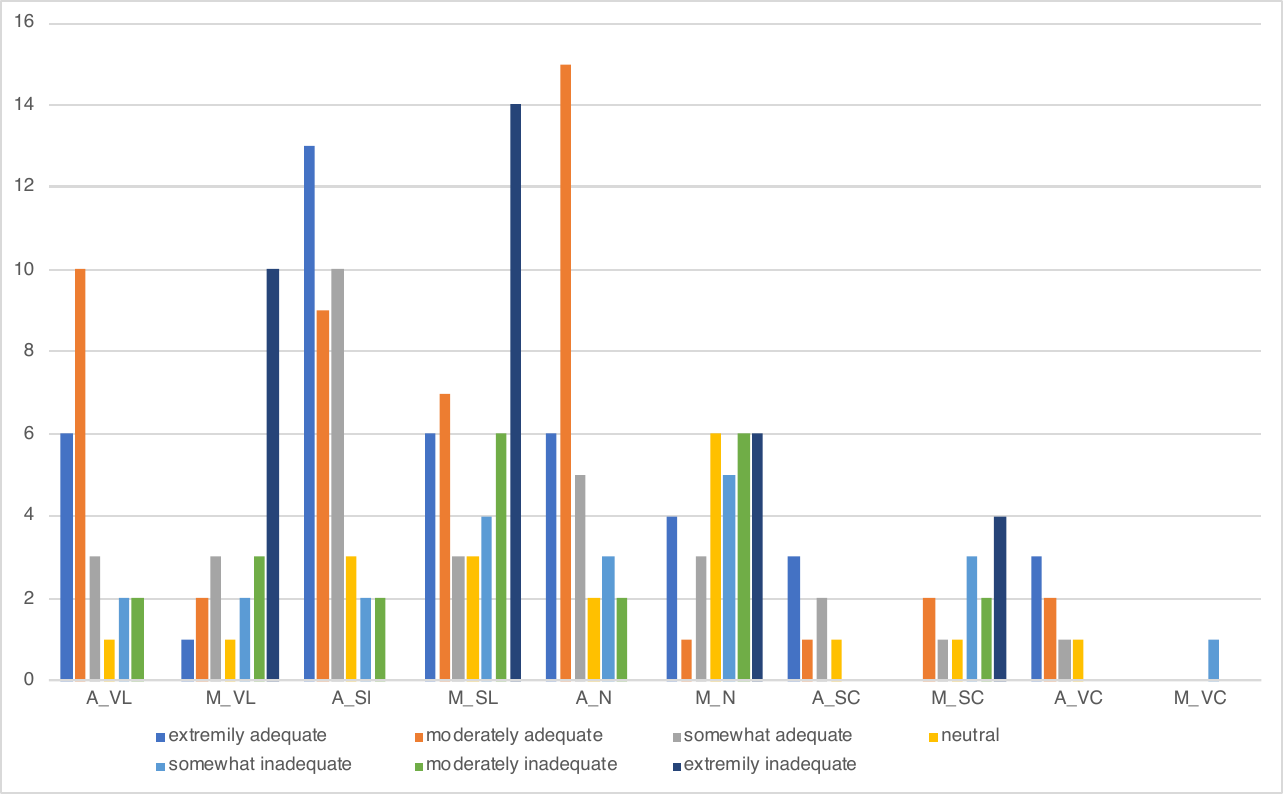}
  \caption{Distribution of Q1 answers per category of political ideology in both Alexa and Malexa Groups.}
  \label{Fig5}
\end{figure}

Figure 5 shows the distribution of responses in both groups on Q1 grouped by the participant's political ideology. The "very liberal" participants are on the opposite ends in the perception of the adequacy of the government response between the Alexa (78.8\% "adequate") and the Malexa groups (68.2\% "inadequate"). The "somewhat liberal" participants in the Alexa group conform with the perception of the "very liberal" participants (82.1\% "adequate") but that is not the case for the "somewhat liberal" participants in the Malexa group which show more uniform distribution in the perception of adequacy. The same conclusion holds for the participants identified as "neither liberal nor conservative" and "somewhat conservative." This finding suggests that Malexa has the potential to utilize the misperception towards creating more divergent views by targeting users outside of the far ends of the political spectrum (due to the liberal-leaning sample, we were unable to compute a meaningful distribution for the "very conservative" category). This is somewhat expected given that the malware used metaphor in the covert manipulation of the OSHA headlines, which is proven to be subjectively experienced and operates on the divergent worldviews of individual participants \cite{Gibbs}.

\begin{figure}[h]
\centering
  \includegraphics[width=\columnwidth]{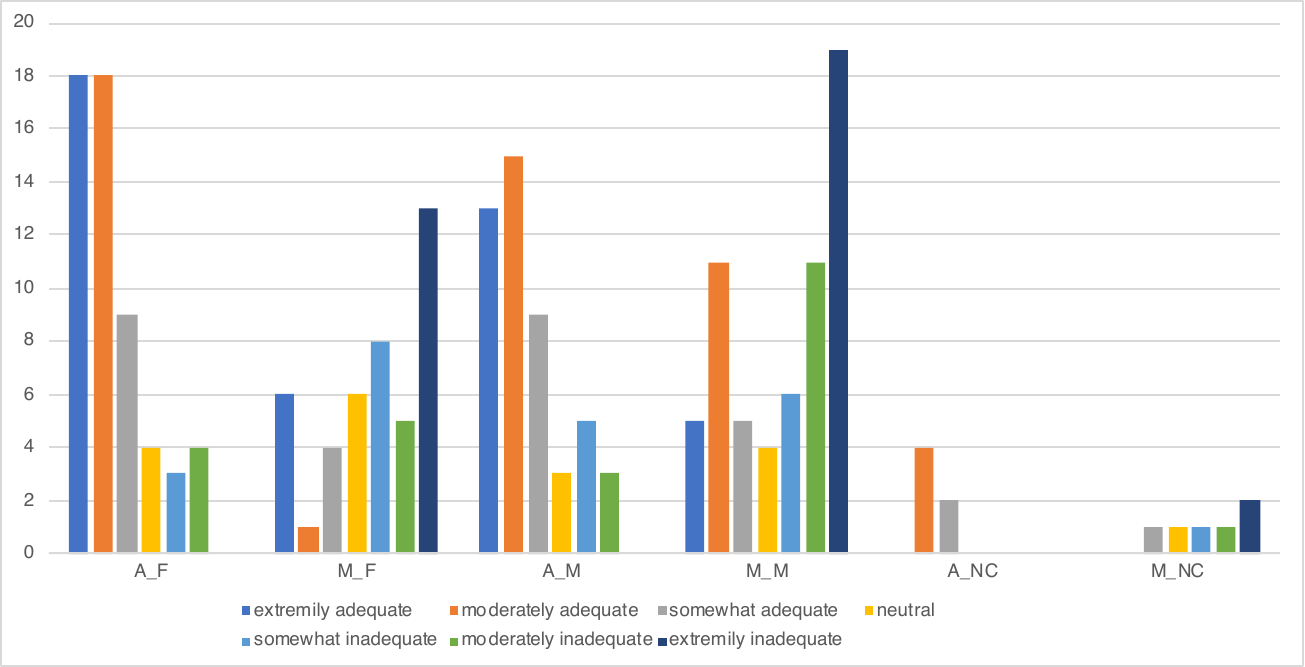}
  \caption{Distribution of Q1 answers per category of gender identity in both Alexa and Malexa Groups.}
  \label{Fig6}
\end{figure}

Figure 6 shows the distribution of responses in both groups on Q1 grouped by the participant's gender identity. More than 80\% of the females, 77\% of the males, and all non-cis participants in the Alexa group perceived the government response as adequate. In the Malexa group, the perception of adequacy drops to 25.6 \% for females, 34.4\% for males, and to 16\% for the non-cis participants. The responses in the Malexa group, when controlled for gender identity, show more uniform distribution in the perception of adequacy of the government response for each group. This finding further reinforces Malexa's potential to induce divergent misperceptions regardless of one's gender identity. In other words, Malexa needs not to worry about the gender of the targeted individual when profiling a potential target for a MIM attack.  

\begin{figure}[h]
\centering
  \includegraphics[width=\columnwidth]{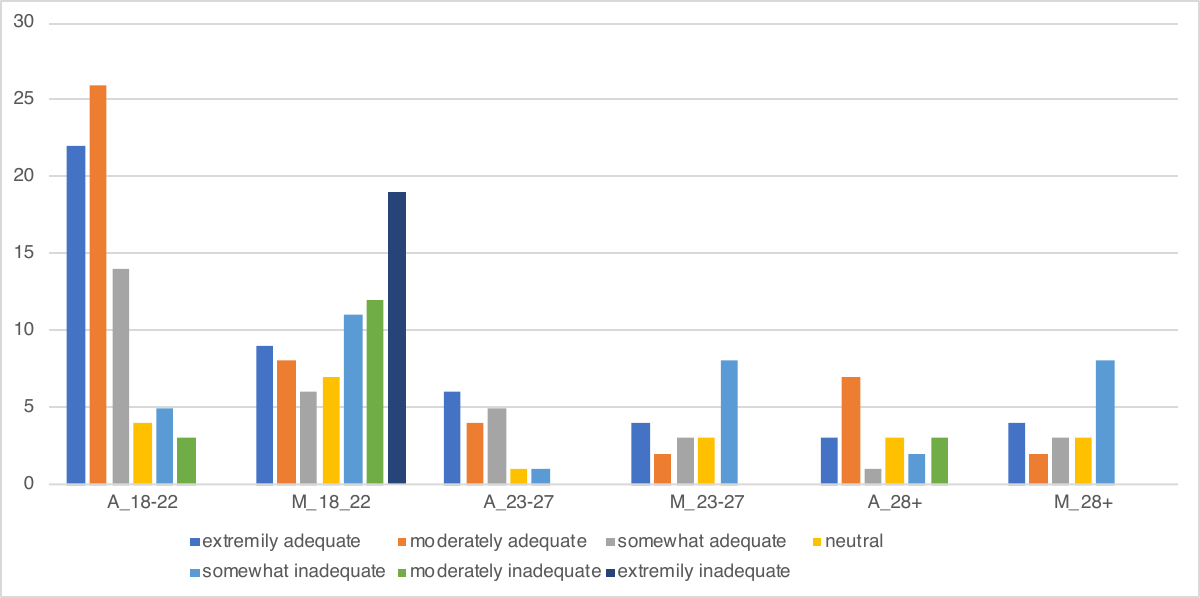}
  \caption{Distribution of Q1 answers per category of gender identity in both Alexa and Malexa Groups.}
  \label{Fig7}
\end{figure}

Figure 7 shows the distribution of responses in both groups on Q1 grouped by the participant's age (categories: 18-22, 23-27, 28+). In the Alexa group, 83.8\% of the participants in the 18-22 age group, 88.2\% in the 23-27 age group, and 57\% in the 28+ age group perceived the government response as adequate. In the Malexa group, the perception of adequacy drops to 31.9\% of the participants in the 18-22 age group, 31.9\% in the 23-27 age group, and 20\% in the 28+ age. Again, the responses in the Malexa group, when controlled for different age groups (in our rather young-leaning sample), show more uniform distribution in the perception of adequacy of the government response for each group. Malexa, according to these findings, is capable of inducing divergent opinion regardless of one's age, too. In other words, Malexa needs not to worry about the age group of the targeted individual when profiling a potential target for a MIM attack.

Hypothesis $1_{2}$ stated that if the workers were the one that benefited in the reported incidents, there won't be any difference in the perception between the two groups (Q2). As shown in Table 1, there is a statistically significant difference in perception between the participants in the Alexa and Malexa groups  $U = 8010.5$, $p < 001$ (effect size = \textit{medium}) that the workers benefited in the reported OSHA incidents. The participants in the Alexa group "moderately agreed" that the government helped the workers while the participants in the Malexa group appear to "slightly disagree." This result rejects Hypothesis $1_{2}$ and accepts the alternative hypothesis, that is, Malexa is capable of inducing misperceptions that the workers didn't benefit from the government response. 

\begin{figure}[h]
\centering
  \includegraphics[width=\columnwidth]{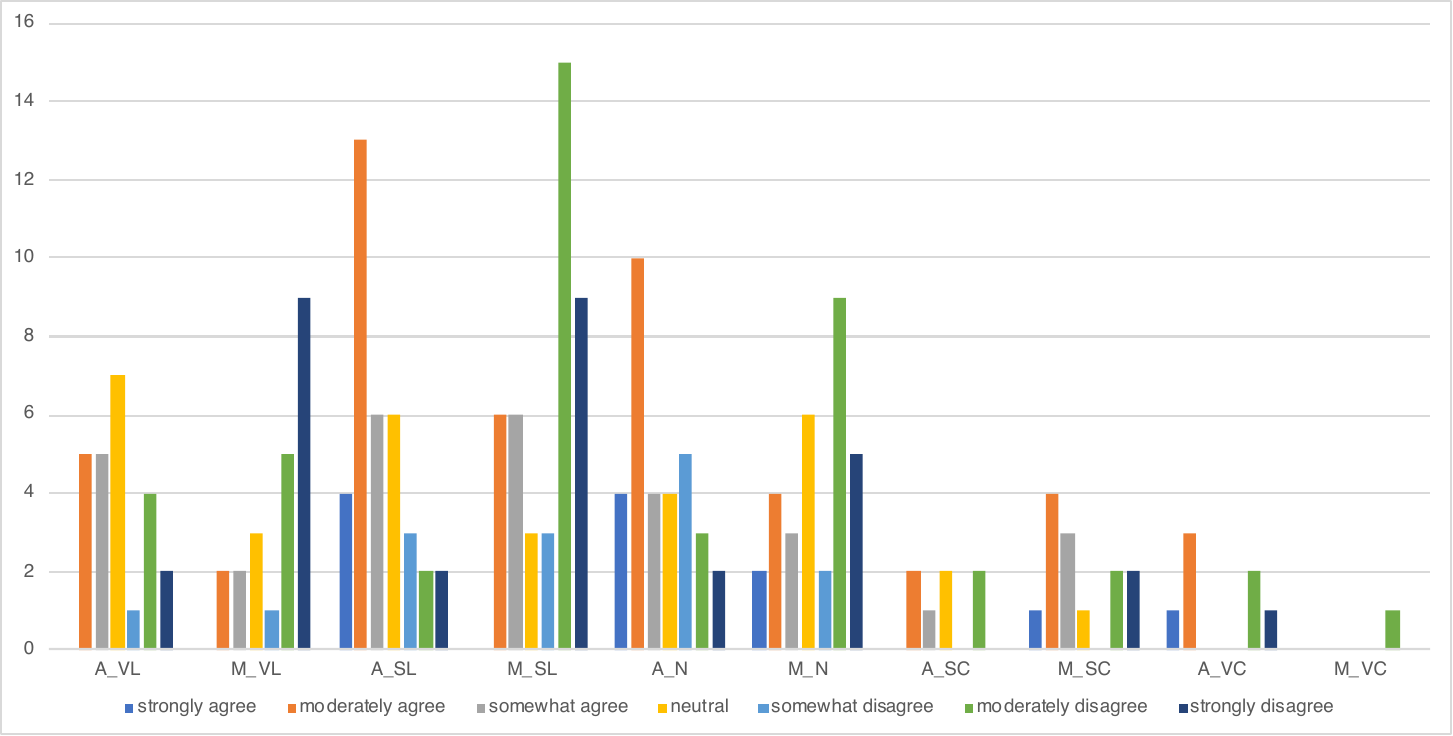}
  \caption{Distribution of Q2 answers per category of political ideology in both Alexa and Malexa Groups.}
  \label{Fig8}
\end{figure}

Figure 8 shows the distribution of responses in both groups on Q2 grouped by the participant's political ideology. The "very liberal" participants in the Alexa group seem more uniformly in the distribution of the responses than the ones in the Malexa groups (68.2\% "disagree"). The "somewhat liberal" participants in the Alexa group (63.9\% "agree") mostly perceived that the workers in the OSHA incidents actually benefited from the government response, while the "somewhat liberal" participants in the Malexa group were more polarized in their perceptions. The "neutral" participants in both groups were more or less polarized in their perceptions on whether the workers in the incident benefited from the government response. The "somewhat conservative" participants in the Alexa group were more uniformly distributed in perception compared to the "somewhat conservative" participants in the Malexa who were more polarized in their perceptions (61.5\% "disagree"). It follows that Malexa can have a polarizing effect for "somewhat liberal" or "somewhat conservative" individuals compared to the far-left or neutral individuals when the government regulation is seen from the perspective of the  workers. 


\begin{figure}[h]
\centering
  \includegraphics[width=\columnwidth]{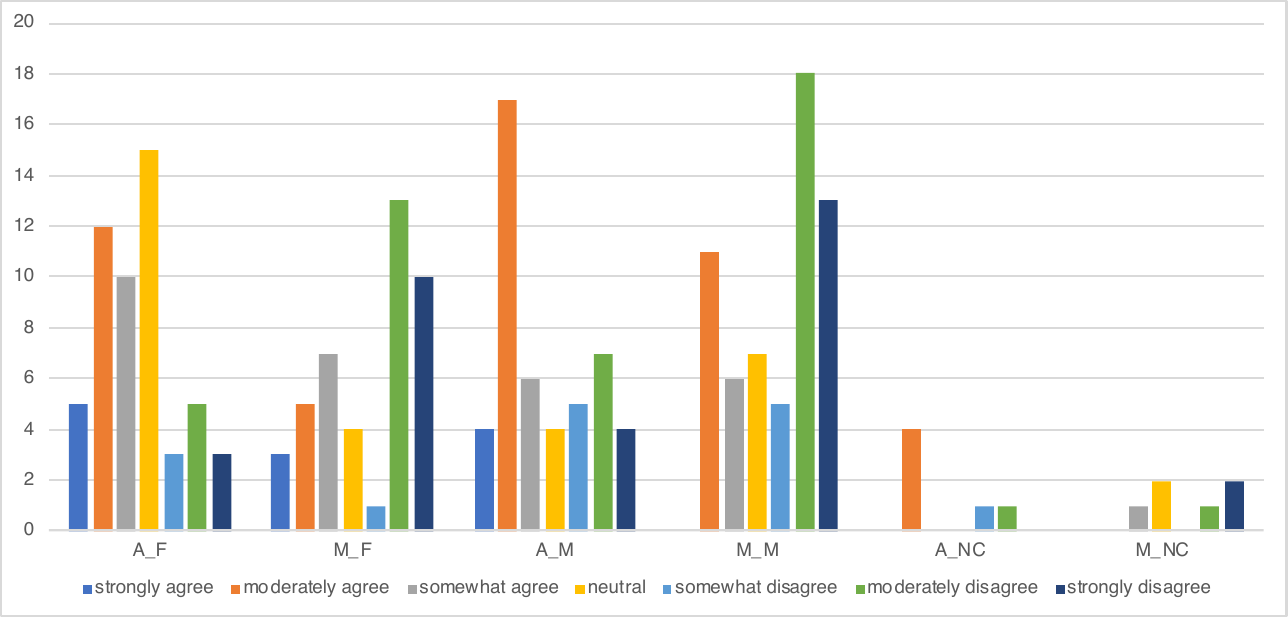}
  \caption{Distribution of Q2 answers per category of gender identity in both Alexa and Malexa Groups.}
  \label{Fig9}
\end{figure}

Figure 9 shows the distribution of responses in both groups on Q2 grouped by the participant's gender identity. More than 50.9\% of the female participants, 57.4\% of the male participants, and 66.7\% of the non-cis participants in the Alexa group perceived that the workers benefited from the government response in the OSHA reports. In the Malexa group, the agreement drops to 34.9 \% for females, 28.3\% for males, and to 16.7\% for the non-cis participants, but the responses appear more polarized (57.3 \% for females, 60\% for males, and to 50\% for the non-cis participants disagreed with the Q2 statement). These results further reinforce the previous finding that Malexa needs not to worry about the gender of the targeted individual when profiling a potential target for a MIM attack. 

\begin{figure}[h]
\centering
  \includegraphics[width=\columnwidth]{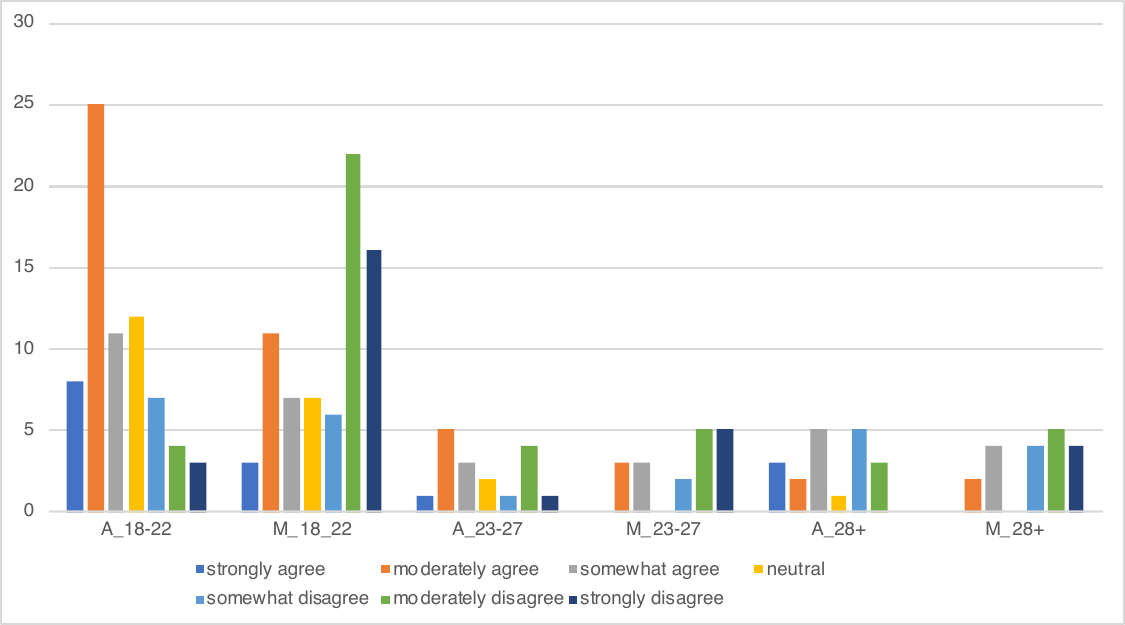}
  \caption{Distribution of Q2 answers per category of gender identity in both Alexa and Malexa Groups.}
  \label{Fig10}
\end{figure}

Figure 10 shows the distribution of responses in both groups on Q2 grouped by the participant's age (categories: 18-22, 23-27, 28+). In the Alexa group, 62.9\% of the participants in the 18-22 age group, 52.9\% in the 23-27 age group, and 52.6\% in the 28+ age group perceived that the workers benefited from the government response in the OSHA reports. In the Malexa group the agreement drops to 29.9\% of the participants in the 18-22 age group, 33.3\% in the 23-27 age group, and 31.6\% in the 28+ age, and following the same polarizing trend as above, the responses appear more polarized (61.1\% of the participants in the 8-22 age group, 55.6\% in the 23-27 age group, and 47.4\% in the 28+ age disagreed with the Q2 statement).  These results further reinforce the previous finding that Malexa needs not to worry about the age group of the targeted individual when profiling a potential target for a MIM attack.

Hypothesis $1_{3}$ stated that if the companies were the one that benefited in the reported incidents, there won't be any difference in the perception between the two groups (Q2). As shown in Table 1, there is a statistically significant difference in perception between the participants in the Alexa and Malexa groups  $U = 3789$, $p < 001$ (effect size = \textit{medium}) that the companies benefited in the reported OSHA incidents. The participants in the Alexa group "moderately disagreed" that the government helped the companies while the participants in the Malexa group "slightly agreed." This result rejects Hypothesis $1_{3}$ and accepts the alternative hypothesis, that is, Malexa is capable of inducing misperceptions that the companies indeed benefited from the government response. 

\begin{figure}[h]
\centering
  \includegraphics[width=\columnwidth]{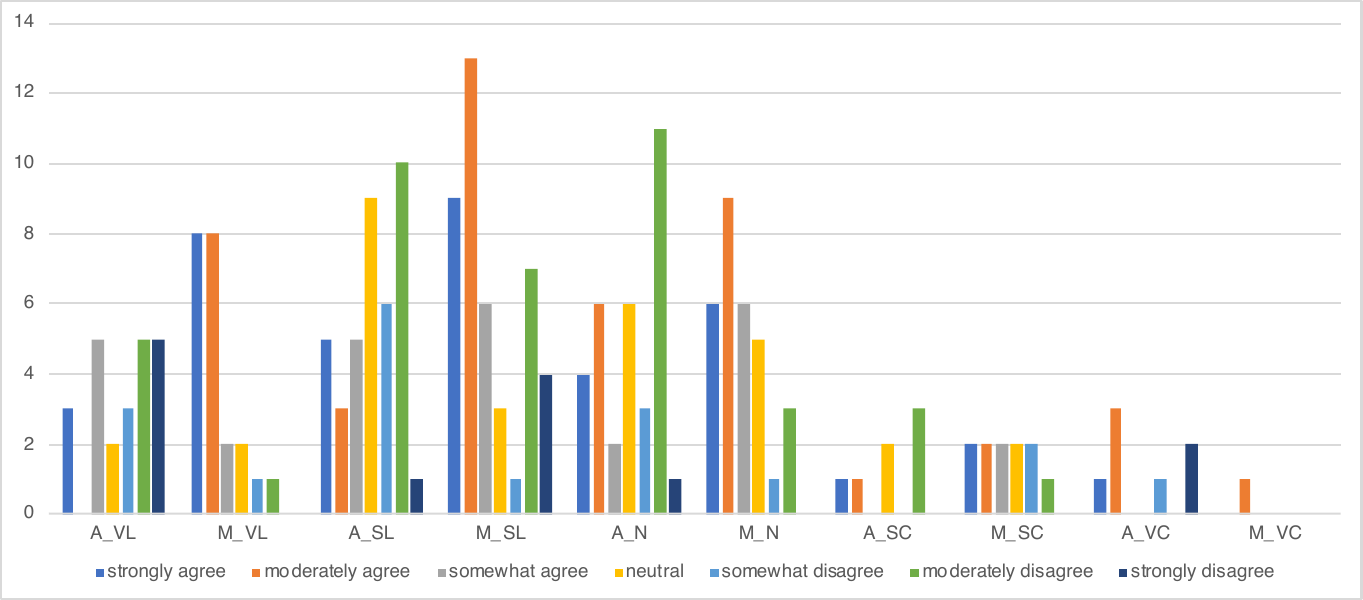}
  \caption{Distribution of Q3 answers per category of political ideology in both Alexa and Malexa Groups.}
  \label{Fig11}
\end{figure}

Figure 11 shows the distribution of responses in both groups on Q3 grouped by the participant's political ideology. The "very liberal" participants in the Alexa group (56.5\% "disagree") seem more polarized compared to the participants in the Malexa groups (81.8\% "agree") who predominately perceived that the companies indeed benefited from the government response. The "somewhat liberal" participants in the Alexa group seem more uniformly distributed in the perceived benefit for the companies in the OSHA reports, while the "somewhat liberal" participants in the Malexa were more polarized in their perceptions (65.1\% "agree"). The "neutral" participants in the Alexa group seem more polarized compared to the participants in the Malexa group (70\% "agree") who predominately perceived that the companies indeed benefited from the government response. The "somewhat conservative" participants in the Alexa group seem more uniformly distributed in the perceived benefit for the companies in the OSHA reports, while the "somewhat conservative" participants in the Malexa group were more polarized in their perceptions (54.5\% "agree"). Similarly as in Q2, these results demonstrating that Malexa can have a polarizing effect for individuals that identify as either "somewhat liberal" or "somewhat conservative" compared to the far-left or neutral individuals when the government regulation is seen from the perspective of the companies involved.

\begin{figure}[h]
\centering
  \includegraphics[width=\columnwidth]{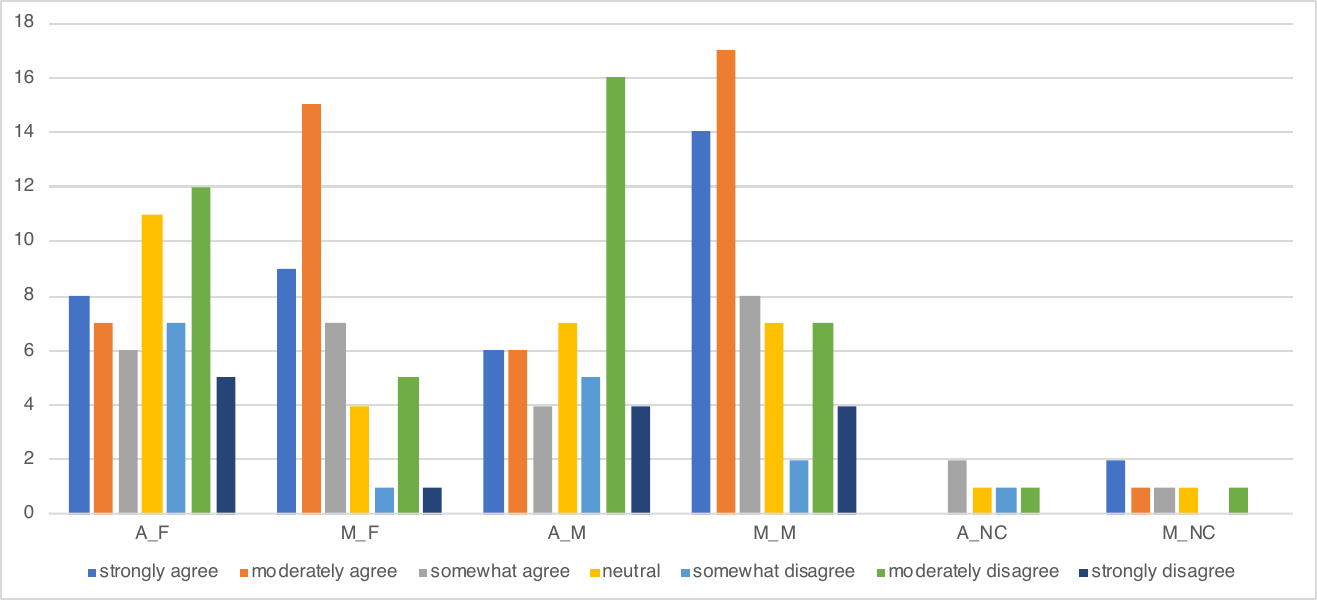}
  \caption{Distribution of Q3 answers per category of gender identity in both Alexa and Malexa Groups.}
  \label{Fig12}
\end{figure}

Figure 12 shows the distribution of responses in both groups on Q3 grouped by the participant's gender identity. In the Alexa group, 37.5\% of the female, 33.3\% of the male, and 40\% of the non-cis participants perceived that the companies instead benefited from the government response in the OSHA reports. In the Malexa group, the agreement rises to 73.8\% for females, 66.1\% for males, and to 66.1\% for the non-cis participants. In this case, it appears that Malexa is capable of reinforcing a particular misperception regardless of one's gender identity. 

\begin{figure}[h]
\centering
  \includegraphics[width=\columnwidth]{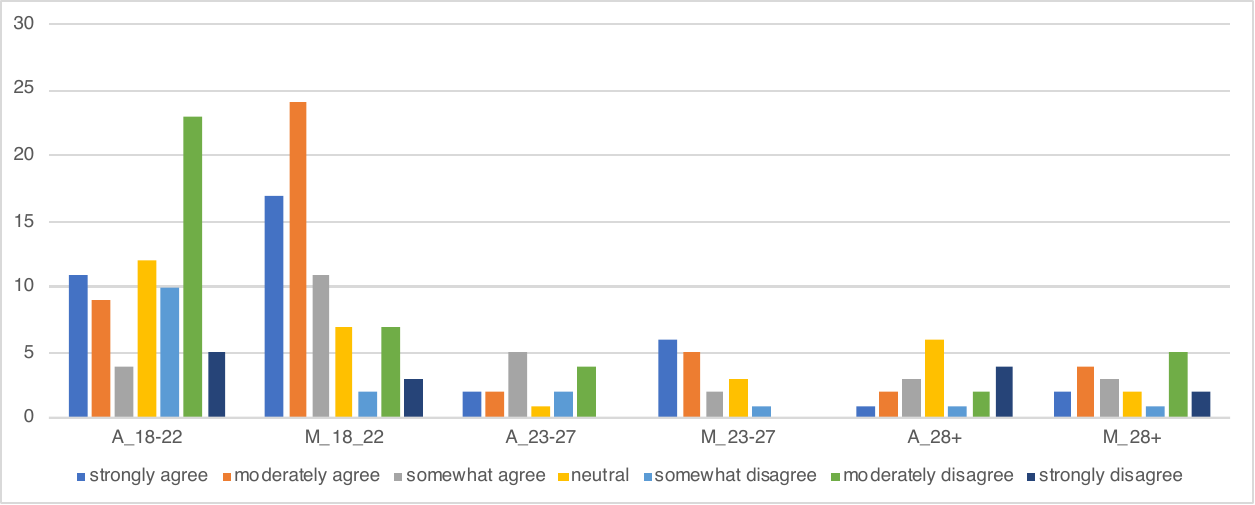}
  \caption{Distribution of Q3 answers per category of gender identity in both Alexa and Malexa Groups.}
  \label{Fig13}
\end{figure}

Figure 13 shows the distribution of responses in both groups on Q3 grouped by the participant's age (categories: 18-22, 23-27, 28+). In the Alexa group, 32.4\% of the participants in the 18-22 age group, 56.3\% in the 23-27 age group, and 31.6\% in the 28+ age group perceived that the workers benefited from the government response in the OSHA reports. In the Malexa group the agreement rises to 73.2\% of the participants in the 18-22 age group, 76.5\% in the 23-27 age group, and 47.4\% in the 28+ age. Similarly, it appears that Malexa is also capable of reinforcing a particular misperception regardless of one's age category. 

In summary, Malexa was found to be capable of inducing misperceptions simply by covertly rewording the headlines and without affecting the factual structure of the news content. This is a somewhat expected result, one might say, given that polarized media outlets resort to similar tactics for pragmatic use of linguistic content formatting (although based on editorial decisions and given an overt political agenda). Therefore we furthered our analysis to explore how Malexa can utilize some public information, like one's political ideology, gender identity, or age. We found that Malexa is capable of inducing misperceptions and creating divergent views by targeting users outside of the far ends of the political spectrum. We also found that Malexa can have a "polarizing"œ and "perception reinforcing" effect regardless of a target user's gender or age.

\subsection{Hypothesis 2}
Hypothesis 2 stated that one's frequency of using intelligent voice assistants doesn't affect their perception of news on the government's regulation performance about workplace safety incidents. We found no statistical significance \textit{within} both the Alexa and Malexa group for the three categories of frequency of intelligent voice assistant use, accepting the Hypothesis 2. This result indicates that the frequency of interaction with intelligent voice assistant doesn't affect the perception formation for target individuals, regardless of the wording of the news spoken back to them. We extended the analysis to compare the perceptions \textit{between} the Alexa and Malexa group when controlling for the frequency of interaction. 

\begin{table}[h]
\renewcommand{\arraystretch}{1.3}
\caption{Q1 perceptions between the Alexa and Malexa groups, controlling for the frequency of interaction.} 
\label{table_2}
\centering
\begin{tabular}{|l|c|c|c|c|}
\hline
& Sometimes & Half time & Most time \\
\hline
\hline
\textbf{Mann-Whitney $U$} & 939.5 & 474 & 145 \\
\hline
\textbf{$Z$ score} & 5.248 & 5.602 & 2.646 \\
\hline
\textbf{Effect size $r$} & .443 & .554 & .390 \\
\hline
\textbf{Asmt. Sig. $p$} & .000* & .000* & .008* \\
\hline
\end{tabular}
\end{table}

As shown in Table 2, there is a statistical significance between the participants' perceptions for all the frequency groups on Q1. The "somewhat" group of participants in the Alexa scenario perceived the government response "somewhat adequate" while the same subgroup in the Malexa group perceived it as "somewhat inadequate." The "half the time" group of participants in the Alexa scenario perceived the government response "somewhat adequate" while the same subgroup in the Malexa group perceived it as "moderately inadequate." The "most of the time" group of participants in the Alexa scenario perceived the government response "somewhat adequate" while the same subgroup in the Malexa group perceived it as "neither adequate nor inadequate."

\begin{table}[!h]
\renewcommand{\arraystretch}{1.3}
\caption{Q2 perceptions between the Alexa and Malexa groups, controlling for the frequency of interaction.} 
\label{table_3}
\centering
\begin{tabular}{|l|c|c|c|c|}
\hline
& Sometimes & Half time & Most time \\
\hline
\hline
\textbf{Mann-Whitney $U$} & 1293.5 & 762 & 173.5 \\
\hline
\textbf{$Z$ score} & 3.728 & 3.396 & 1.602 \\
\hline
\textbf{Effect size $r$} & .305 & .339 & .160 \\
\hline
\textbf{Asmt. Sig. $p$} & .000* & .001* & .108 \\
\hline
\end{tabular}
\end{table}

As shown in Table 3, there is a statistical significance between the participants' perceptions on Q2 only for the "sometimes" and "half the time" groups. The "somewhat" group of participants in the Alexa scenario "slightly agreed" that the workers benefited from the government response, while they "somewhat disagreed" in the Malexa scenario. The "half the time" group of participants in the Alexa scenario "neither agreed nor disagreed" that the workers benefited from the government response, while they "moderately disagreed" in the Malexa scenario.

\begin{table}[h]
\renewcommand{\arraystretch}{1.3}
\caption{Q3 perceptions between the Alexa and Malexa groups, controlling for the frequency of interaction.} 
\label{table_s}
\centering
\begin{tabular}{|l|c|c|c|c|}
\hline
& Sometimes & Half time & Most time \\
\hline
\hline
\textbf{Mann-Whitney $U$} & 1706 & 1766 & 272.5 \\
\hline
\textbf{$Z$ score} & 2.825 & 3.4 & 1.056 \\
\hline
\textbf{Effect size $r$} & .332 & .338 & .161 \\
\hline
\textbf{Asmt. Sig. $p$} & .005* & .001* & .291 \\
\hline
\end{tabular}
\end{table}

As shown in Table 4, there is a statistical significance between the participants' perceptions on Q3 only for the "sometimes" and "half the time" groups. Both the "sometimes" and "half the time" participants in the Alexa scenario "neither agreed nor disagreed" that the companies benefited from the government response, while they "somewhat agreed" in the Malexa scenario. In summary, the results show that Malexa can break the ambivalence for people who interact with intelligent voice assistants either sometimes or half the time. These findings indicate that a malicious actor can profile the target users not just by political ideology, but also by frequency of interaction. Following the principle of moderation, here too, the most likely targets of the MIM attack are those that supplement their daily news diet with information from intelligent voice assistants, but not entirely rely on them. 




\section{Discussion}
\subsection{What Malexa Did in Our Study}
This study tested the effects of a malicious third-party skill aiming to covertly manipulate the perception about workplace safety news delivered by an intelligent voice assistant to a targeted user. To do so, the malicious skill covertly launches a MIM attack on the content spoken back to a targeted user, turning Alexa into Malexa. Previous studies exploring malicious third-party skills manipulated the invocation name of the skill with the goal to impersonate an original skill. Although this certainly works in our case and may even be a favorable way of skill invocation, our primary interest was to test whether the victims of the MIM attack in an intelligent voice assistant ecosystem will report different perceptions than users receiving original, non-manipulated content. 


The results demonstrate that Malexa is capable of manipulating the perception of the government action in the reported OSHA news. Malexa essentially altered the perspective in these OSHA news releases (or "the cover story" as referred to in the misperception chain) and with that made the participants react as if the government inadequately favored the big business instead of the workers in the safety violations (the opposite reaction was observed for the participants in the Alexa group). We are aware that "adequate" as a response is a subjective valuation respective to the context of the reported incident, one's trust in government, or a particular regulatory awareness. We see the statistically significant difference in the reported perception between groups as a preliminary evidence of the Malexa's potential as an "influencer" and not definitive proof that Malexa can do actually "brainwashing" \cite{Aro}.  

The potential for "influencing" is further corroborated with the results suggesting that one's political ideology could be used to manipulate their perceptions when talking to Alexa or Malexa. This gives Malexa the opportunity to go beyond the political influencing Cambridge Analytica did during the 2016 election cycle \cite{Baldwin}. Instead of playing with inflammatory content, Malexa's strategy is to "nudge" a targeted user (either "somewhat liberal" or "somewhat conservative") to subjectively assess the news headlines. That is, Malexa has the advantage to reframe a political argument to appeal to one's partisanship when they don't stand on the far ends of the political spectrum \cite{Day}. Malexa also has the advantage in picking up the influencing targets based on how frequently a user interacts with an intelligent voice assistant. The results from our study suggest that the most attractive targets, at least in the context of government regulating news, were the ones using voice assistants moderately, that is, more for getting quick information rather as a main source of information. 

\subsection{What Malexa Can't Do Now}
As much as we have extolled the "virtues" of Malexa, the MIM attack in an intelligent voice assistant ecosystem has a long way to go. In our study, Malexa was preconfigured to do the replacements for the particular valence word/phrase pairs and affect only the perception on government regulation when it comes to briefings on workplace safety violations. Even if the rewording logic can be updated dynamically, it takes time for a MIM attacker to create "the cover story," e.g. decide what valence word/phrase pairs to use and when to time the particular rewording. This is not a trivial part of the attack. Neither is the execution, given that the attacker needs to get the targeted user to download the Malexa skill in the first place (alternatively, use a squatting/command hijacking attack).    

We chose to work with OSHA news releases and measured the perceptions only on three arbitrary selected news headlines. If Malexa worked with, say breaking news headlines or more/less number of headlines, the results might be different than what we observed in our study. This is something we are highly interested in and we are preparing a next round of a Malexa study to comparatively assess its MIM effectiveness based on various types (amount) of stories it voices back to users, e.g. politics, sport, local news, foreign affairs, and entertainment. We are also interested to see how Malexa can affect perception of social media content.

Malexa, in the presented form, might have a problem with scaling and remaining undetected by a wide intelligent voice assistant user base. Users might hear the headlines from Alexa but their primary interface for online information is usually a computer, web browser, or an app. For example, a targeted user can hear what Malexa has to say but read a news article covering the same incident and notice that something is wrong. We took a precaution to make sure that Malexa changes the user/s perception but doesn't distort the main facts in the news headlines we used in our study. Nonetheless, it is sufficient for a targeted user to simply deem that the third-party skill "acts weird" or that "something phishy" is going on and remove it from their device. Caution is warranted when interpreting the results of our study. We used a convenience sample of intelligent voice assistant users skewed towards a younger, urban, liberal leaning population and future studies might investigate and yield different results based on these demographic factors. 

\subsection{What Malexa Could Do in the Future}
Despite the obvious limitations, the findings of this study provide reasonable ground to discuss how Malexa could be weaponized in the future. We already mentioned the potential of Malexa becoming a covert "influencer," a cyberoperations vector interesting for actors wishing to interfere with major events like elections \cite{Bradshaw}. So far, actors resorted to trolling campaigns targeting specific populations on controversial topics discussed via social media \cite{DiResta}. However, this requires generating a lot of \textit{fabricated} content in order to be successful \cite{Paavola}. Malexa requires no such thing because it operates on legitimate content. Malexa, in a much more targeted fashion, can determine what a target will hear back (or not, Malexa can be configured to even suppress certain news headlines if they are deemed unfavorable to the objective of the deception campaign). In our descriptive MIM attack example, Malexa might omit the part of the headline about Senator Sanders's future presidential campaign plans and only report the part about his heart attack. 

Another advantage of Malexa is that it can tailor the MIM attack to different news and valence word/phrase pairs to target different users because the delivery of information is interpersonal rather than public as in the trolling campaigns \cite{Cowan}. What Alice heard about Senator Sanders might not be what Bob did, and if they don't particularly talk about it, they are both deceived to believe different "cover stories." The best advantage of Malexa is, perhaps, the fact that we openly welcome "her" in our homes \cite{Kiseleva}. Most of the discussions on intelligent voice assistants revolve around what they "listen to" but not what they "say." Facebook trolling posts could be missed, omitted, detected, or ignored because they are visually inspected. It is harder to do so with Malexa because of the voice interface and the interpersonal nature of interaction \cite{Purington}. Who will think that Malexa, sitting quietly on the nightstand and ordering our usual groceries from Whole Foods, might be manipulating us behind the scenes ~\cite {Laubheimer}? 

\subsection{How to Counter Malexa}
Malexa relies on an attacker registering a seemingly legitimate skill. The certification process should look for any news delivery logic that dynamically rearranges words outside of the content pulled from an external RSS feed or URL. It should potentially require the skill to work with RSS feeds or URLs that allow for content integrity verification by the user on the Alexa phone application. The Alexa smartphone app can follow a similar example to Facebook to increase the transparency of the third-party behind a skill. Facebook, aiming in part to combat election interference, includes a new "Organizations That Manage This Page" tab that shows an organizations' legal name and verified city, phone number or website \cite{facebook}. One thing to have in mind is the possibility of a MIM attacker trying to sneak Malexa as an "accessibility (a11y) skill," claiming that the rewording is done to create an assistive natural language skill that for example helps non-native English speakers \cite{Jang}. It might be harder to bar a skill from the Alexa Skills Store on these grounds, therefore, the certification process must request all the use cases for these word manipulations upfront to ensure no MIM is hidden in the content delivery logic of the assistive application.

It's worth nothing that the MIM attacker has the alternative to share the skill with other users via direct link, bypassing the Alexa Skills Store. In this case, we recommend that the user verifies the publisher (e.g. "CIA" and not \textit{Laynr} as with the "CIA Factbook" skill) and more importantly, the delivery logic. If a skill claims it delivers news briefings from say, The New York Times, it is fairly easy for a user to run a quick test and compare the headlines with the online version "verbatim." However, users should be aware that even legitimate news sources sometimes use different headlines depending on the interface (e.g. a phone app or a web browser) or the edition \cite{Bulik}. This is done because interfaces might require different limits on the number of characters in the headline. Or, to increase readership for an initially poor performing headlines by rewording them with clear, powerful words and a conversational tone (interestingly, a MIM attacker can use similar claims to justify that the rewording makes the news "sound better when spoken back"). This may affect the point of using a news skill and can lead to false positives, but carefully looking for deception cues might help users spot and remove a malicious skill \cite{Nicholson}. 

The ethical implications of Malexa are the same as those related to publishing any vulnerability: the value of publicly sharing a proof-of-concept malicious skill with knowledgeable researchers outweighs the opportunity that potential attackers may benefit from the publication. If this paper introduces a viable attack in the intelligent voice assistant ecosystem--which we believe it will--due to the simplistic nature of this attack, we believe adversaries will develop and deploy similar attacks, if they have not already. The study itself tests the plausibility of Malexa with a proof-of-concept, locally-executed skill extension (not publicly available on the Alexa Skills Store). In the context of a real-life Malexa, a responsible disclosure would entail contacting Amazon and working with them through the details of the MIM attack. 


\section{Conclusion}
In this work, we introduced Malexa, Alexa's malicious twin. Malexa's goal is to covertly manipulate the user's perception on the content delivered by an intelligent voice assistant. We showed how an attacker, implementing the malware-induced misperception attack, could turn Alexa into Malexa without the user's knowledge. We then tested Malexa in a study with 220 participants asking for their perception on the government's regulation performance based on OSHA news briefings. Our findings show that Malexa is able to make the government look much less friendly to the working people and more in favor of big businesses. Malexa showed that "she" is able to manipulate perceptions regardless of one's political ideology or how frequently one uses a voice assistant. Unlike visually inspected text of Facebook or Twitter posts, text-to-speech voice provides little cues for hearing inspection concerning whether trolling content is used or not. People readily welcome and undoubtedly trust intelligent voice assistants and let them sit in their homes. While people are concerned about what Alexa "listens to," little do they pay attention to what is actually "spoken back to" them. If Facebook and Twitter are gearing up to combat trolling in the next election cycle, Malexa, as demonstrated in our study, provides a new avenue for delivering trolling content directly into our homes. We hope our results inform the security community about the implications of having an alternative vector for any micro-targeted influence.


\bibliographystyle{ACM-Reference-Format}
\bibliography{mim}


\begin{thebibliography}{56}


\ifx \showCODEN    \undefined \def \showCODEN     #1{\unskip}     \fi
\ifx \showDOI      \undefined \def \showDOI       #1{#1}\fi
\ifx \showISBNx    \undefined \def \showISBNx     #1{\unskip}     \fi
\ifx \showISBNxiii \undefined \def \showISBNxiii  #1{\unskip}     \fi
\ifx \showISSN     \undefined \def \showISSN      #1{\unskip}     \fi
\ifx \showLCCN     \undefined \def \showLCCN      #1{\unskip}     \fi
\ifx \shownote     \undefined \def \shownote      #1{#1}          \fi
\ifx \showarticletitle \undefined \def \showarticletitle #1{#1}   \fi
\ifx \showURL      \undefined \def \showURL       {\relax}        \fi
\providecommand\bibfield[2]{#2}
\providecommand\bibinfo[2]{#2}
\providecommand\natexlab[1]{#1}
\providecommand\showeprint[2][]{arXiv:#2}

\bibitem[\protect\citeauthoryear{Amazon}{Amazon}{2019a}]%
        {Amazon}
\bibfield{author}{\bibinfo{person}{Amazon}.} \bibinfo{year}{2019}\natexlab{a}.
\newblock \bibinfo{title}{{Alexa Skills Kit}}.
\newblock
\newblock
\newblock
\shownote{\url {https://developer.amazon.com/en-US/alexa/alexa-skills-kit}.}


\bibitem[\protect\citeauthoryear{Amazon}{Amazon}{2019b}]%
        {alexa_skill_blueprints}
\bibfield{author}{\bibinfo{person}{Amazon}.} \bibinfo{year}{2019}\natexlab{b}.
\newblock \bibinfo{title}{{Amazon Alexa | Skill Blueprints}}.
\newblock
\newblock
\newblock
\shownote{\url{https://blueprints.amazon.com/}.}


\bibitem[\protect\citeauthoryear{Ammari, Kaye, Tsai, and Bentley}{Ammari
  et~al\mbox{.}}{2019}]%
        {Ammari}
\bibfield{author}{\bibinfo{person}{Tawfiq Ammari}, \bibinfo{person}{Jofish
  Kaye}, \bibinfo{person}{Janice~Y. Tsai}, {and} \bibinfo{person}{Frank
  Bentley}.} \bibinfo{year}{2019}\natexlab{}.
\newblock \showarticletitle{{Music, Search, and IoT: How People (Really) Use
  Voice Assistants}}.
\newblock \bibinfo{journal}{\emph{ACM Trans. Comput.-Hum. Interact.}}
  \bibinfo{volume}{26}, \bibinfo{number}{3}, Article \bibinfo{articleno}{17}
  (\bibinfo{date}{April} \bibinfo{year}{2019}), \bibinfo{numpages}{28}~pages.
\newblock
\showISSN{1073-0516}
\urldef\tempurl%
\url{https://doi.org/10.1145/3311956}
\showDOI{\tempurl}


\bibitem[\protect\citeauthoryear{Anderson}{Anderson}{2017}]%
        {Anderson}
\bibfield{author}{\bibinfo{person}{Mae Anderson}.}
  \bibinfo{year}{2017}\natexlab{}.
\newblock \bibinfo{title}{{How Burger King revealed the hackability of voice
  assistants}}.
\newblock
\newblock
\newblock
\shownote{\url
  {https://phys.org/news/2017-05-burger-king-revealed-hackability-voice.html}.}


\bibitem[\protect\citeauthoryear{Aro}{Aro}{2016}]%
        {Aro}
\bibfield{author}{\bibinfo{person}{Jessikka Aro}.}
  \bibinfo{year}{2016}\natexlab{}.
\newblock \showarticletitle{{The Cyberspace War: Propaganda and Trolling as
  Warfare Tools}}.
\newblock \bibinfo{journal}{\emph{European View}} \bibinfo{volume}{15},
  \bibinfo{number}{1} (\bibinfo{date}{2019/11/05} \bibinfo{year}{2016}),
  \bibinfo{pages}{121--132}.
\newblock
\showISBNx{1781-6858}
\urldef\tempurl%
\url{https://doi.org/10.1007/s12290-016-0395-5}
\showDOI{\tempurl}


\bibitem[\protect\citeauthoryear{Bakshy, Messing, and Adamic}{Bakshy
  et~al\mbox{.}}{2015}]%
        {Bakshy}
\bibfield{author}{\bibinfo{person}{Eytan Bakshy}, \bibinfo{person}{Solomon
  Messing}, {and} \bibinfo{person}{Lada~A. Adamic}.}
  \bibinfo{year}{2015}\natexlab{}.
\newblock \showarticletitle{{Exposure to ideologically diverse news and opinion
  on Facebook}}.
\newblock \bibinfo{journal}{\emph{Science}} \bibinfo{volume}{348},
  \bibinfo{number}{6239} (\bibinfo{year}{2015}), \bibinfo{pages}{1130--1132}.
\newblock
\showISSN{0036-8075}
\urldef\tempurl%
\url{https://doi.org/10.1126/science.aaa1160}
\showDOI{\tempurl}


\bibitem[\protect\citeauthoryear{Baldwin}{Baldwin}{2018}]%
        {Baldwin}
\bibfield{author}{\bibinfo{person}{Tom Baldwin}.}
  \bibinfo{year}{2018}\natexlab{}.
\newblock \bibinfo{booktitle}{\emph{{Ctrl Alt Delete: How Politics and the
  Media Crashed our Democracy}}}.
\newblock \bibinfo{publisher}{Oxford University Press},
  \bibinfo{address}{Oxford, UK}. 383 pages.
\newblock


\bibitem[\protect\citeauthoryear{Benkler, Faris, and Roberts}{Benkler
  et~al\mbox{.}}{2018}]%
        {Benkler}
\bibfield{author}{\bibinfo{person}{Yochai Benkler}, \bibinfo{person}{Robert
  Faris}, {and} \bibinfo{person}{Hal Roberts}.}
  \bibinfo{year}{2018}\natexlab{}.
\newblock \bibinfo{booktitle}{\emph{{Network Propaganda: Manipulation,
  Disinformation, and Radicalization in American Politics}}}.
\newblock \bibinfo{publisher}{Oxford University Press},
  \bibinfo{address}{Oxford, UK}. 394 pages.
\newblock


\bibitem[\protect\citeauthoryear{Bradshaw and Howard}{Bradshaw and
  Howard}{2017}]%
        {Bradshaw}
\bibfield{author}{\bibinfo{person}{Samantha Bradshaw} {and}
  \bibinfo{person}{Philip~N. Howard}.} \bibinfo{year}{2017}\natexlab{}.
\newblock \bibinfo{booktitle}{\emph{{Troops, Trolls and Troublemakers: A Global
  Inventory of Organized Social Media Manipulation}}}.
\newblock \bibinfo{type}{Technical Report}. \bibinfo{institution}{Oxford
  University, Project on Computational Propaganda}, \bibinfo{address}{Oxford,
  UK}.
\newblock


\bibitem[\protect\citeauthoryear{Carlini, Mishra, Vaidya, Zhang, Sherr,
  Shields, Wagner, and Zhou}{Carlini et~al\mbox{.}}{2016}]%
        {Carlini}
\bibfield{author}{\bibinfo{person}{Nicholas Carlini}, \bibinfo{person}{Pratyush
  Mishra}, \bibinfo{person}{Tavish Vaidya}, \bibinfo{person}{Yuankai Zhang},
  \bibinfo{person}{Micah Sherr}, \bibinfo{person}{Clay Shields},
  \bibinfo{person}{David Wagner}, {and} \bibinfo{person}{Wenchao Zhou}.}
  \bibinfo{year}{2016}\natexlab{}.
\newblock \showarticletitle{{Hidden Voice Commands}}. In
  \bibinfo{booktitle}{\emph{25th {USENIX} Security Symposium ({USENIX} Security
  16)}}. \bibinfo{publisher}{{USENIX} Association}, \bibinfo{address}{Austin,
  TX}, \bibinfo{pages}{513--530}.
\newblock
\showISBNx{978-1-931971-32-4}
\urldef\tempurl%
\url{https://www.usenix.org/conference/usenixsecurity16/technical-sessions/presentation/carlini}
\showURL{%
\tempurl}


\bibitem[\protect\citeauthoryear{{Chung}, {Iorga}, {Voas}, and {Lee}}{{Chung}
  et~al\mbox{.}}{2017}]%
        {Chung}
\bibfield{author}{\bibinfo{person}{H. {Chung}}, \bibinfo{person}{M. {Iorga}},
  \bibinfo{person}{J. {Voas}}, {and} \bibinfo{person}{S. {Lee}}.}
  \bibinfo{year}{2017}\natexlab{}.
\newblock \showarticletitle{{Alexa, Can I Trust You?}}
\newblock \bibinfo{journal}{\emph{Computer}} \bibinfo{volume}{50},
  \bibinfo{number}{9} (\bibinfo{year}{2017}), \bibinfo{pages}{100--104}.
\newblock
\urldef\tempurl%
\url{https://doi.org/10.1109/MC.2017.3571053}
\showDOI{\tempurl}


\bibitem[\protect\citeauthoryear{Cowan, Pantidi, Coyle, Morrissey, Clarke,
  Al-Shehri, Earley, and Bandeira}{Cowan et~al\mbox{.}}{2017}]%
        {Cowan}
\bibfield{author}{\bibinfo{person}{Benjamin~R. Cowan}, \bibinfo{person}{Nadia
  Pantidi}, \bibinfo{person}{David Coyle}, \bibinfo{person}{Kellie Morrissey},
  \bibinfo{person}{Peter Clarke}, \bibinfo{person}{Sara Al-Shehri},
  \bibinfo{person}{David Earley}, {and} \bibinfo{person}{Natasha Bandeira}.}
  \bibinfo{year}{2017}\natexlab{}.
\newblock \showarticletitle{{"What Can I Help You with?": Infrequent Users'
  Experiences of Intelligent Personal Assistants}}. In
  \bibinfo{booktitle}{\emph{Proceedings of the 19th International Conference on
  Human-Computer Interaction with Mobile Devices and Services}}
  \emph{(\bibinfo{series}{MobileHCI '17})}. \bibinfo{publisher}{ACM},
  \bibinfo{address}{New York, NY, USA}, Article \bibinfo{articleno}{43},
  \bibinfo{numpages}{12}~pages.
\newblock
\showISBNx{978-1-4503-5075-4}
\urldef\tempurl%
\url{https://doi.org/10.1145/3098279.3098539}
\showDOI{\tempurl}


\bibitem[\protect\citeauthoryear{Day, Fiske, Downing, and Trail}{Day
  et~al\mbox{.}}{2014}]%
        {Day}
\bibfield{author}{\bibinfo{person}{Martin~V. Day}, \bibinfo{person}{Susan~T.
  Fiske}, \bibinfo{person}{Emily~L. Downing}, {and} \bibinfo{person}{Thomas~E.
  Trail}.} \bibinfo{year}{2014}\natexlab{}.
\newblock \showarticletitle{{Shifting Liberal and Conservative Attitudes Using
  Moral Foundations Theory}}.
\newblock \bibinfo{journal}{\emph{Personality and Social Psychology Bulletin}}
  \bibinfo{volume}{40}, \bibinfo{number}{12} (\bibinfo{date}{2019/11/05}
  \bibinfo{year}{2014}), \bibinfo{pages}{1559--1573}.
\newblock
\showISBNx{0146-1672}
\urldef\tempurl%
\url{https://doi.org/10.1177/0146167214551152}
\showDOI{\tempurl}


\bibitem[\protect\citeauthoryear{Diao, Liu, Zhou, and Zhang}{Diao
  et~al\mbox{.}}{2014}]%
        {Diao}
\bibfield{author}{\bibinfo{person}{Wenrui Diao}, \bibinfo{person}{Xiangyu Liu},
  \bibinfo{person}{Zhe Zhou}, {and} \bibinfo{person}{Kehuan Zhang}.}
  \bibinfo{year}{2014}\natexlab{}.
\newblock \showarticletitle{Your Voice Assistant is Mine: How to Abuse Speakers
  to Steal Information and Control Your Phone}. In
  \bibinfo{booktitle}{\emph{Proceedings of the 4th ACM Workshop on Security and
  Privacy in Smartphones \&\#38; Mobile Devices}} \emph{(\bibinfo{series}{SPSM
  '14})}. \bibinfo{publisher}{ACM}, \bibinfo{address}{New York, NY, USA},
  \bibinfo{pages}{63--74}.
\newblock
\showISBNx{978-1-4503-3155-5}
\urldef\tempurl%
\url{https://doi.org/10.1145/2666620.2666623}
\showDOI{\tempurl}


\bibitem[\protect\citeauthoryear{DiResta, Shaffer, Ruppel, Becky, Sullivan,
  David, Matney, Fox, Albright, and Johnson}{DiResta et~al\mbox{.}}{2018}]%
        {DiResta}
\bibfield{author}{\bibinfo{person}{Renee DiResta}, \bibinfo{person}{Kris
  Shaffer}, \bibinfo{person}{Ruppel}, \bibinfo{person}{Becky},
  \bibinfo{person}{Sullivan}, \bibinfo{person}{David}, \bibinfo{person}{Robert
  Matney}, \bibinfo{person}{Ryan Fox}, \bibinfo{person}{Jonathan Albright},
  {and} \bibinfo{person}{Ben Johnson}.} \bibinfo{year}{2018}\natexlab{}.
\newblock \bibinfo{booktitle}{\emph{{The Tactics and Tropes of the Internet
  Research Agency}}}.
\newblock \bibinfo{type}{Technical Report}. \bibinfo{institution}{New
  Knowledge}.
\newblock


\bibitem[\protect\citeauthoryear{Ember and Martin}{Ember and Martin}{2019}]%
        {Ember}
\bibfield{author}{\bibinfo{person}{Sydney Ember} {and}
  \bibinfo{person}{Jonathan Martin}.} \bibinfo{year}{2019}\natexlab{}.
\newblock \bibinfo{title}{{Bernie Sanders Says He Will Slow His Campaign Pace
  After Heart Attack}}.
\newblock
\newblock
\newblock
\shownote{\url
  {https://www.nytimes.com/2019/10/08/us/politics/bernie-sanders-heart-attack.html?searchResultPosition=7}.}


\bibitem[\protect\citeauthoryear{Facebook}{Facebook}{2019}]%
        {facebook}
\bibfield{author}{\bibinfo{person}{Facebook}.} \bibinfo{year}{2019}\natexlab{}.
\newblock \bibinfo{title}{{Helping to Protect the 2020 US Elections}}.
\newblock
\newblock
\newblock
\shownote{\url{https://newsroom.fb.com/news/2019/10/update-on-election-integrity-efforts/}.}


\bibitem[\protect\citeauthoryear{Foley}{Foley}{2019}]%
        {kmeans_text_blog_2}
\bibfield{author}{\bibinfo{person}{Daniel Foley}.}
  \bibinfo{year}{2019}\natexlab{}.
\newblock \bibinfo{title}{{K-Means Clustering: Making Sense of Text Data using
  Unsupervised Learning}}.
\newblock
\newblock
\newblock
\shownote{\url
  {https://towardsdatascience.com/k-means-clustering-8e1e64c1561c}.}


\bibitem[\protect\citeauthoryear{for Economic Co-operation and (OECD)}{for
  Economic Co-operation and (OECD)}{2012}]%
        {oecd}
\bibfield{author}{\bibinfo{person}{Organization for Economic Co-operation}
  {and} \bibinfo{person}{Development (OECD)}.} \bibinfo{year}{2012}\natexlab{}.
\newblock \bibinfo{title}{{Measuring Regulatory Performance: A Practitioner's
  Guide to Perception Surveys}}.
\newblock
\newblock
\urldef\tempurl%
\url{https://doi.org/10.1787/9789264167179-en}
\showDOI{\tempurl}
\newblock
\shownote{\url {https://www.oecd.org/gov/regulatory-policy/48933826.pdf}.}


\bibitem[\protect\citeauthoryear{Gibbs~Jr}{Gibbs~Jr}{2008}]%
        {Gibbs}
\bibfield{author}{\bibinfo{person}{Raymond~W Gibbs~Jr}.}
  \bibinfo{year}{2008}\natexlab{}.
\newblock \bibinfo{booktitle}{\emph{The Cambridge handbook of metaphor and
  thought}}.
\newblock \bibinfo{publisher}{Cambridge University Press},
  \bibinfo{address}{Cambridge, UK}.
\newblock


\bibitem[\protect\citeauthoryear{Goel}{Goel}{2018}]%
        {kmeans_text_blog_3}
\bibfield{author}{\bibinfo{person}{Vishabh Goel}.}
  \bibinfo{year}{2018}\natexlab{}.
\newblock \bibinfo{title}{{Applying Machine Learning to Classify an
  Unsupervised Text Document}}.
\newblock
\newblock
\newblock
\shownote{\url
  {https://towardsdatascience.com/applying-machine-learning-to-classify-an-unsupervised-text-document-e7bb6265f52}.}


\bibitem[\protect\citeauthoryear{Granville}{Granville}{2018}]%
        {Granville}
\bibfield{author}{\bibinfo{person}{K Granville}.}
  \bibinfo{year}{2018}\natexlab{}.
\newblock \bibinfo{title}{{Facebook and Cambridge Analytica: What You Need to
  Know as Fallout Widens}}.
\newblock
\newblock
\urldef\tempurl%
\url{https://www.nytimes.com/2018/03/19/technology/facebook-cambridge-analytica-explained.html}
\showURL{%
\tempurl}


\bibitem[\protect\citeauthoryear{{Heckman}, {Stech}, {Schmoker}, and
  {Thomas}}{{Heckman} et~al\mbox{.}}{2015}]%
        {Heckman}
\bibfield{author}{\bibinfo{person}{K.~E. {Heckman}}, \bibinfo{person}{F.~J.
  {Stech}}, \bibinfo{person}{B.~S. {Schmoker}}, {and} \bibinfo{person}{R.~K.
  {Thomas}}.} \bibinfo{year}{2015}\natexlab{}.
\newblock \showarticletitle{{Denial and Deception in Cyber Defense}}.
\newblock \bibinfo{journal}{\emph{Computer}} \bibinfo{volume}{48},
  \bibinfo{number}{4} (\bibinfo{date}{Apr} \bibinfo{year}{2015}),
  \bibinfo{pages}{36--44}.
\newblock
\urldef\tempurl%
\url{https://doi.org/10.1109/MC.2015.104}
\showDOI{\tempurl}


\bibitem[\protect\citeauthoryear{Initiative}{Initiative}{2012}]%
        {NIST}
\bibfield{author}{\bibinfo{person}{Joint Task Force~Transformation
  Initiative}.} \bibinfo{year}{2012}\natexlab{}.
\newblock \bibinfo{booktitle}{\emph{Guide for Conducting Risk Assessments}}.
\newblock \bibinfo{type}{Technical Reportt} 800-30.
  \bibinfo{institution}{National Institute of Standards and Technology},
  \bibinfo{address}{Gaithersburg, MD}.
\newblock


\bibitem[\protect\citeauthoryear{Jang, Song, Chung, Wang, and Lee}{Jang
  et~al\mbox{.}}{2014}]%
        {Jang}
\bibfield{author}{\bibinfo{person}{Yeongjin Jang}, \bibinfo{person}{Chengyu
  Song}, \bibinfo{person}{Simon~P. Chung}, \bibinfo{person}{Tielei Wang}, {and}
  \bibinfo{person}{Wenke Lee}.} \bibinfo{year}{2014}\natexlab{}.
\newblock \showarticletitle{{A11Y Attacks: Exploiting Accessibility in
  Operating Systems}}. In \bibinfo{booktitle}{\emph{Proceedings of the 2014 ACM
  SIGSAC Conference on Computer and Communications Security}}
  \emph{(\bibinfo{series}{CCS '14})}. \bibinfo{publisher}{ACM},
  \bibinfo{address}{New York, NY, USA}, \bibinfo{pages}{103--115}.
\newblock
\showISBNx{978-1-4503-2957-6}
\urldef\tempurl%
\url{https://doi.org/10.1145/2660267.2660295}
\showDOI{\tempurl}


\bibitem[\protect\citeauthoryear{Kiseleva, Williams, Jiang, Hassan~Awadallah,
  Crook, Zitouni, and Anastasakos}{Kiseleva et~al\mbox{.}}{2016}]%
        {Kiseleva}
\bibfield{author}{\bibinfo{person}{Julia Kiseleva}, \bibinfo{person}{Kyle
  Williams}, \bibinfo{person}{Jiepu Jiang}, \bibinfo{person}{Ahmed
  Hassan~Awadallah}, \bibinfo{person}{Aidan~C. Crook}, \bibinfo{person}{Imed
  Zitouni}, {and} \bibinfo{person}{Tasos Anastasakos}.}
  \bibinfo{year}{2016}\natexlab{}.
\newblock \showarticletitle{Understanding User Satisfaction with Intelligent
  Assistants}. In \bibinfo{booktitle}{\emph{Proceedings of the 2016 ACM on
  Conference on Human Information Interaction and Retrieval}}
  \emph{(\bibinfo{series}{CHIIR '16})}. \bibinfo{publisher}{ACM},
  \bibinfo{address}{New York, NY, USA}, \bibinfo{pages}{121--130}.
\newblock
\showISBNx{978-1-4503-3751-9}
\urldef\tempurl%
\url{https://doi.org/10.1145/2854946.2854961}
\showDOI{\tempurl}


\bibitem[\protect\citeauthoryear{Kumar, Paccagnella, Murley, Hennenfent, Mason,
  Bates, and Bailey}{Kumar et~al\mbox{.}}{2018}]%
        {Kumar}
\bibfield{author}{\bibinfo{person}{Deepak Kumar}, \bibinfo{person}{Riccardo
  Paccagnella}, \bibinfo{person}{Paul Murley}, \bibinfo{person}{Eric
  Hennenfent}, \bibinfo{person}{Joshua Mason}, \bibinfo{person}{Adam Bates},
  {and} \bibinfo{person}{Michael Bailey}.} \bibinfo{year}{2018}\natexlab{}.
\newblock \showarticletitle{{Skill Squatting Attacks on Amazon Alexa}}. In
  \bibinfo{booktitle}{\emph{27th {USENIX} Security Symposium ({USENIX} Security
  18)}}. \bibinfo{publisher}{{USENIX} Association},
  \bibinfo{address}{Baltimore, MD}, \bibinfo{pages}{33--47}.
\newblock
\showISBNx{978-1-939133-04-5}
\urldef\tempurl%
\url{https://www.usenix.org/conference/usenixsecurity18/presentation/kumar}
\showURL{%
\tempurl}


\bibitem[\protect\citeauthoryear{Lau, Zimmerman, and Schaub}{Lau
  et~al\mbox{.}}{2018}]%
        {Lau}
\bibfield{author}{\bibinfo{person}{Josephine Lau}, \bibinfo{person}{Benjamin
  Zimmerman}, {and} \bibinfo{person}{Florian Schaub}.}
  \bibinfo{year}{2018}\natexlab{}.
\newblock \showarticletitle{{Alexa, Are You Listening?: Privacy Perceptions,
  Concerns and Privacy-seeking Behaviors with Smart Speakers}}.
\newblock \bibinfo{journal}{\emph{Proc. ACM Hum.-Comput. Interact.}}
  \bibinfo{volume}{2}, \bibinfo{number}{CSCW}, Article \bibinfo{articleno}{102}
  (\bibinfo{date}{Nov.} \bibinfo{year}{2018}), \bibinfo{numpages}{31}~pages.
\newblock
\showISSN{2573-0142}
\urldef\tempurl%
\url{https://doi.org/10.1145/3274371}
\showDOI{\tempurl}


\bibitem[\protect\citeauthoryear{Laubheimer and Budiu}{Laubheimer and
  Budiu}{2018}]%
        {Laubheimer}
\bibfield{author}{\bibinfo{person}{Page Laubheimer} {and}
  \bibinfo{person}{Raluca Budiu}.} \bibinfo{year}{2018}\natexlab{}.
\newblock \bibinfo{title}{{Intelligent Assistants: Creepy, Childish, or a Tool?
  Users’ Attitudes Toward Alexa, Google Assistant, and Siri}}.
\newblock
\newblock
\newblock
\shownote{\url{https://www.nngroup.com/articles/voice-assistant-attitudes/}.}


\bibitem[\protect\citeauthoryear{Laynr}{Laynr}{2017}]%
        {world_factbook}
\bibfield{author}{\bibinfo{person}{Laynr}.} \bibinfo{year}{2017}\natexlab{}.
\newblock \bibinfo{title}{Applying Machine Learning to Classify an Unsupervised
  Text Document}.
\newblock
\newblock
\newblock
\shownote{\url{https://www.amazon.com/Laynr-World-Factbook/dp/B01DMQ0QN4/ref=sr_1_2?keywords=factbook&qid=1572835240&s=digital-skills&sr=1-2}.}


\bibitem[\protect\citeauthoryear{Levine}{Levine}{2014}]%
        {Levine}
\bibfield{author}{\bibinfo{person}{Timothy~R Levine}.}
  \bibinfo{year}{2014}\natexlab{}.
\newblock \showarticletitle{{Truth-Default Theory ({TDT})}}.
\newblock \bibinfo{journal}{\emph{Journal of Language and Social Psychology}}
  \bibinfo{volume}{33}, \bibinfo{number}{4} (\bibinfo{year}{2014}),
  \bibinfo{pages}{378--392}.
\newblock


\bibitem[\protect\citeauthoryear{Lu, Qin, Cao, Liu, and Mengxing}{Lu
  et~al\mbox{.}}{2014}]%
        {kmeans}
\bibfield{author}{\bibinfo{person}{Meilian Lu}, \bibinfo{person}{Zhen Qin},
  \bibinfo{person}{Yiming Cao}, \bibinfo{person}{Zhichao Liu}, {and}
  \bibinfo{person}{Wang Mengxing}.} \bibinfo{year}{2014}\natexlab{}.
\newblock \showarticletitle{{Scalable News Recommendation using
  Multi-Dimensional Similarity and Jaccard–Kmeans Clustering}}.
\newblock \bibinfo{journal}{\emph{Journal of Systems and Software}}
  \bibinfo{volume}{95} (\bibinfo{year}{2014}), \bibinfo{pages}{242--251}.
\newblock


\bibitem[\protect\citeauthoryear{Mark}{Mark}{2016}]%
        {Bulik}
\bibfield{author}{\bibinfo{person}{Bulik Mark}.}
  \bibinfo{year}{2016}\natexlab{}.
\newblock \bibinfo{title}{{Which Headlines Attract Most Readers?}}
\newblock
\newblock
\newblock
\shownote{\url{https://www.nytimes.com/2016/06/13/insider/which-headlines-attract-most-readers.html
  }.}


\bibitem[\protect\citeauthoryear{Newman}{Newman}{2018}]%
        {Newman}
\bibfield{author}{\bibinfo{person}{Lily~Hay Newman}.}
  \bibinfo{year}{2018}\natexlab{}.
\newblock \bibinfo{title}{{Chrome Extension Malware Has Evolved}}.
\newblock
\newblock
\newblock
\shownote{\url {https://www.wired.com/story/chrome-extension-malware/}.}


\bibitem[\protect\citeauthoryear{Nicholson, Coventry, and Briggs}{Nicholson
  et~al\mbox{.}}{2017}]%
        {Nicholson}
\bibfield{author}{\bibinfo{person}{James Nicholson}, \bibinfo{person}{Lynne
  Coventry}, {and} \bibinfo{person}{Pam Briggs}.}
  \bibinfo{year}{2017}\natexlab{}.
\newblock \showarticletitle{{Can we fight social engineering attacks by social
  means? Assessing social salience as a means to improve phish detection}}. In
  \bibinfo{booktitle}{\emph{Thirteenth Symposium on Usable Privacy and Security
  ({SOUPS} 2017)}}. \bibinfo{publisher}{{USENIX} Association},
  \bibinfo{address}{Santa Clara, CA}, \bibinfo{pages}{285--298}.
\newblock


\bibitem[\protect\citeauthoryear{Paavola, Helo, Jalonen, Sartonen, and
  Huhtinen}{Paavola et~al\mbox{.}}{2016}]%
        {Paavola}
\bibfield{author}{\bibinfo{person}{J Paavola}, \bibinfo{person}{T Helo},
  \bibinfo{person}{H Jalonen}, \bibinfo{person}{M Sartonen}, {and}
  \bibinfo{person}{A-M Huhtinen}.} \bibinfo{year}{2016}\natexlab{}.
\newblock \showarticletitle{{Understanding the Trolling Phenomenon}}.
\newblock \bibinfo{journal}{\emph{Journal of Information Warfare}}
  \bibinfo{volume}{15}, \bibinfo{number}{4} (\bibinfo{year}{2016}),
  \bibinfo{pages}{100--111}.
\newblock


\bibitem[\protect\citeauthoryear{Parkin, Pater, Lopez-Neira, and
  Tanczer}{Parkin et~al\mbox{.}}{2019}]%
        {Parkin}
\bibfield{author}{\bibinfo{person}{Simon Parkin}, \bibinfo{person}{Trupti
  Pater}, \bibinfo{person}{Isabel Lopez-Neira}, {and} \bibinfo{person}{Leonie
  Tanczer}.} \bibinfo{year}{2019}\natexlab{}.
\newblock \showarticletitle{{Usability Analysis of Shared Device Ecosystem
  Security: Informing Support for Survivors of IoT-Facilitated Tech-Abuse}}. In
  \bibinfo{booktitle}{\emph{New Security Paradigms Workshop ({NSPW} 2019)}}.
  \bibinfo{publisher}{ACM}, \bibinfo{address}{San Carlos, Costa Rica},
  \bibinfo{pages}{to appear}.
\newblock


\bibitem[\protect\citeauthoryear{Pfeifle}{Pfeifle}{2018}]%
        {Pfeifle}
\bibfield{author}{\bibinfo{person}{Anne Pfeifle}.}
  \bibinfo{year}{2018}\natexlab{}.
\newblock \showarticletitle{{Alexa, What Should We Do about Privacy: Protecting
  Privacy for Users of Voice-Activated Devices Comments}}.
\newblock \bibinfo{journal}{\emph{Washington Law Review}} \bibinfo{volume}{93},
  \bibinfo{number}{1} (\bibinfo{year}{2018}), \bibinfo{pages}{421--458}.
\newblock


\bibitem[\protect\citeauthoryear{Purington, Taft, Sannon, Bazarova, and
  Taylor}{Purington et~al\mbox{.}}{2017}]%
        {Purington}
\bibfield{author}{\bibinfo{person}{Amanda Purington},
  \bibinfo{person}{Jessie~G. Taft}, \bibinfo{person}{Shruti Sannon},
  \bibinfo{person}{Natalya~N. Bazarova}, {and} \bibinfo{person}{Samuel~Hardman
  Taylor}.} \bibinfo{year}{2017}\natexlab{}.
\newblock \showarticletitle{{"Alexa is My New BFF": Social Roles, User
  Satisfaction, and Personification of the Amazon Echo}}. In
  \bibinfo{booktitle}{\emph{Proceedings of the 2017 CHI Conference Extended
  Abstracts on Human Factors in Computing Systems}} \emph{(\bibinfo{series}{CHI
  EA '17})}. \bibinfo{publisher}{ACM}, \bibinfo{address}{New York, NY, USA},
  \bibinfo{pages}{2853--2859}.
\newblock
\showISBNx{978-1-4503-4656-6}
\urldef\tempurl%
\url{https://doi.org/10.1145/3027063.3053246}
\showDOI{\tempurl}


\bibitem[\protect\citeauthoryear{Rodionova}{Rodionova}{2019}]%
        {Rodionova}
\bibfield{author}{\bibinfo{person}{Zlata Rodionova}.}
  \bibinfo{year}{2019}\natexlab{}.
\newblock \bibinfo{title}{Burger King ad backfires after asking Google what's
  in a Whopper and is told 'cyanide'}.
\newblock
\newblock
\newblock
\shownote{\url{https://www.independent.co.uk/news/business/news/burger-king-advert-ask-google-big-whopper-cyanide-cancer-causing-wikipedia-page-us-a7681561.html}.}


\bibitem[\protect\citeauthoryear{Roy, Shen, Hassanieh, and Choudhury}{Roy
  et~al\mbox{.}}{2018}]%
        {Roy}
\bibfield{author}{\bibinfo{person}{Nirupam Roy}, \bibinfo{person}{Sheng Shen},
  \bibinfo{person}{Haitham Hassanieh}, {and} \bibinfo{person}{Romit~Roy
  Choudhury}.} \bibinfo{year}{2018}\natexlab{}.
\newblock \showarticletitle{Inaudible Voice Commands: The Long-Range Attack and
  Defense}. In \bibinfo{booktitle}{\emph{15th {USENIX} Symposium on Networked
  Systems Design and Implementation ({NSDI} 18)}}. \bibinfo{publisher}{{USENIX}
  Association}, \bibinfo{address}{Renton, WA}, \bibinfo{pages}{547--560}.
\newblock
\showISBNx{978-1-939133-01-4}
\urldef\tempurl%
\url{https://www.usenix.org/conference/nsdi18/presentation/roy}
\showURL{%
\tempurl}


\bibitem[\protect\citeauthoryear{Russell}{Russell}{2013}]%
        {Russell}
\bibfield{author}{\bibinfo{person}{Matthew~A. Russell}.}
  \bibinfo{year}{2013}\natexlab{}.
\newblock \bibinfo{booktitle}{\emph{{Mining the Social Web: Data Mining
  Facebook, Twitter, LinkedIn, Google+, GitHub, and More}}}.
\newblock \bibinfo{publisher}{O'Reilly Media}.
\newblock


\bibitem[\protect\citeauthoryear{Salnikov}{Salnikov}{2018}]%
        {kmeans_text_blog_1}
\bibfield{author}{\bibinfo{person}{Mikhail Salnikov}.}
  \bibinfo{year}{2018}\natexlab{}.
\newblock \bibinfo{title}{Text Clustering with K-Means and TF-IDF}.
\newblock
\newblock
\newblock
\shownote{\url{https://medium.com/@MSalnikov/text-clustering-with-k-means-and-tf-idf-f099bcf95183}.}


\bibitem[\protect\citeauthoryear{Sanders}{Sanders}{2019}]%
        {SandersOSHA}
\bibfield{author}{\bibinfo{person}{Bernie Sanders}.}
  \bibinfo{year}{2019}\natexlab{}.
\newblock \bibinfo{title}{{OSHA Letter Amazon}}.
\newblock
\newblock
\newblock
\shownote{\url{https://www.sanders.senate.gov/download/osha-letter-amazon?id=434F26EA-04F3-44B0-86EC-1B696EBAA2E0&download=1&inline=file}.}


\bibitem[\protect\citeauthoryear{Shirani-Mehr, Rothschild, Goel, and
  Gelman}{Shirani-Mehr et~al\mbox{.}}{2018}]%
        {Mehr}
\bibfield{author}{\bibinfo{person}{Houshmand Shirani-Mehr},
  \bibinfo{person}{David Rothschild}, \bibinfo{person}{Sharad Goel}, {and}
  \bibinfo{person}{Andrew Gelman}.} \bibinfo{year}{2018}\natexlab{}.
\newblock \showarticletitle{{Disentangling Bias and Variance in Election
  Polls}}.
\newblock \bibinfo{journal}{\emph{J. Amer. Statist. Assoc.}}
  \bibinfo{volume}{113}, \bibinfo{number}{522} (\bibinfo{year}{2018}),
  \bibinfo{pages}{607--614}.
\newblock
\urldef\tempurl%
\url{https://doi.org/10.1080/01621459.2018.1448823}
\showDOI{\tempurl}


\bibitem[\protect\citeauthoryear{Simonite}{Simonite}{2019}]%
        {Simonite}
\bibfield{author}{\bibinfo{person}{Tom Simonite}.}
  \bibinfo{year}{2019}\natexlab{}.
\newblock \bibinfo{title}{{Google May Have Finally Made a Truly Usable Voice
  Assistant}}.
\newblock
\newblock
\newblock
\shownote{\url
  {https://www.wired.com/story/google-made-truly-usable-voice-assistant/}.}


\bibitem[\protect\citeauthoryear{Singh, Tiwari, and Garg}{Singh
  et~al\mbox{.}}{2011}]%
        {best_kmeans}
\bibfield{author}{\bibinfo{person}{Vivek~Kumar Singh}, \bibinfo{person}{Nisha
  Tiwari}, {and} \bibinfo{person}{Shekhar Garg}.}
  \bibinfo{year}{2011}\natexlab{}.
\newblock \showarticletitle{{Document Clustering using K-means, Heuristic
  K-means and Fuzzy C-means}}. In \bibinfo{booktitle}{\emph{2011 International
  Conference on Computational Intelligence and Communication Networks}}.
  \bibinfo{publisher}{IEEE}, \bibinfo{address}{Gwalior, India},
  \bibinfo{pages}{7--9}.
\newblock
\showISBNx{978-0-7695-4587-5}
\urldef\tempurl%
\url{https://ieeexplore.ieee.org/document/6112875}
\showURL{%
\tempurl}


\bibitem[\protect\citeauthoryear{Sopory}{Sopory}{2017}]%
        {Sopory}
\bibfield{author}{\bibinfo{person}{Pradeep Sopory}.}
  \bibinfo{year}{2017}\natexlab{}.
\newblock \bibinfo{booktitle}{\emph{{Metaphor in Health and Risk
  Communication}}}.
\newblock \bibinfo{publisher}{Oxford University Press}.
\newblock


\bibitem[\protect\citeauthoryear{Spring}{Spring}{2019}]%
        {Spring}
\bibfield{author}{\bibinfo{person}{Tom Spring}.}
  \bibinfo{year}{2019}\natexlab{}.
\newblock \bibinfo{title}{{DEF CON 2019: Researchers Demo Hacking Google Home
  for RCE}}.
\newblock
\newblock
\newblock
\shownote{\url
  {https://threatpost.com/def-con-2019-hacking-google-home/147170/}.}


\bibitem[\protect\citeauthoryear{{US Department of Labor}}{{US Department of
  Labor}}{2019a}]%
        {dol-osha}
\bibfield{author}{\bibinfo{person}{{US Department of Labor}}.}
  \bibinfo{year}{2019}\natexlab{a}.
\newblock \bibinfo{title}{{Occupational Safety and Hazard Administration News
  Releases}}.
\newblock
\newblock
\newblock
\shownote{\url {https://www.osha.gov/news/newsreleases/}.}


\bibitem[\protect\citeauthoryear{{US Department of Labor}}{{US Department of
  Labor}}{2019b}]%
        {osha-rss}
\bibfield{author}{\bibinfo{person}{{US Department of Labor}}.}
  \bibinfo{year}{2019}\natexlab{b}.
\newblock \bibinfo{title}{{RSS Feeds: Occupational Safety and Hazard
  Administration (OSHA)}}.
\newblock
\newblock
\newblock
\shownote{\url{https://www.osha.gov/rss/index.html}.}


\bibitem[\protect\citeauthoryear{Vaidya, Zhang, Sherr, and Shields}{Vaidya
  et~al\mbox{.}}{2015}]%
        {Tavish}
\bibfield{author}{\bibinfo{person}{Tavish Vaidya}, \bibinfo{person}{Yuankai
  Zhang}, \bibinfo{person}{Micah Sherr}, {and} \bibinfo{person}{Clay Shields}.}
  \bibinfo{year}{2015}\natexlab{}.
\newblock \showarticletitle{{Cocaine Noodles: Exploiting the Gap between Human
  and Machine Speech Recognition}}. In \bibinfo{booktitle}{\emph{9th {USENIX}
  Workshop on Offensive Technologies ({WOOT} 15)}}.
  \bibinfo{publisher}{{USENIX} Association}, \bibinfo{address}{Washington,
  D.C.}
\newblock
\urldef\tempurl%
\url{https://www.usenix.org/conference/woot15/workshop-program/presentation/vaidya}
\showURL{%
\tempurl}


\bibitem[\protect\citeauthoryear{Vincent}{Vincent}{2018}]%
        {Vincent}
\bibfield{author}{\bibinfo{person}{James Vincent}.}
  \bibinfo{year}{2018}\natexlab{}.
\newblock \bibinfo{title}{{This blessed Chrome extension replaces 'Elon Musk'
  with 'Grimes's Boyfriend'}}.
\newblock
\newblock
\newblock
\shownote{\url
  {https://www.theverge.com/tldr/2018/5/10/17338984/elon-musk-grimes-boyfriend-chrome-extension}.}


\bibitem[\protect\citeauthoryear{Zeng, Mare, and Roesner}{Zeng
  et~al\mbox{.}}{2017}]%
        {Zeng}
\bibfield{author}{\bibinfo{person}{Eric Zeng}, \bibinfo{person}{Shrirang Mare},
  {and} \bibinfo{person}{Franziska Roesner}.} \bibinfo{year}{2017}\natexlab{}.
\newblock \showarticletitle{{End User Security and Privacy Concerns with Smart
  Homes}}. In \bibinfo{booktitle}{\emph{Thirteenth Symposium on Usable Privacy
  and Security ({SOUPS} 2017)}}. \bibinfo{publisher}{{USENIX} Association},
  \bibinfo{address}{Santa Clara, CA}, \bibinfo{pages}{65--80}.
\newblock
\showISBNx{978-1-931971-39-3}
\urldef\tempurl%
\url{https://www.usenix.org/conference/soups2017/technical-sessions/presentation/zeng}
\showURL{%
\tempurl}


\bibitem[\protect\citeauthoryear{{Zhang}, {Mi}, {Feng}, {Wang}, {Tian}, and
  {Qian}}{{Zhang} et~al\mbox{.}}{2019}]%
        {Zhang}
\bibfield{author}{\bibinfo{person}{N. {Zhang}}, \bibinfo{person}{X. {Mi}},
  \bibinfo{person}{X. {Feng}}, \bibinfo{person}{X. {Wang}}, \bibinfo{person}{Y.
  {Tian}}, {and} \bibinfo{person}{F. {Qian}}.} \bibinfo{year}{2019}\natexlab{}.
\newblock \showarticletitle{{Dangerous Skills: Understanding and Mitigating
  Security Risks of Voice-Controlled Third-Party Functions on Virtual Personal
  Assistant Systems}}. In \bibinfo{booktitle}{\emph{2019 IEEE Symposium on
  Security and Privacy (SP)}}. \bibinfo{pages}{1381--1396}.
\newblock
\urldef\tempurl%
\url{https://doi.org/10.1109/SP.2019.00016}
\showDOI{\tempurl}


\bibitem[\protect\citeauthoryear{Zheng, Apthorpe, Chetty, and Feamster}{Zheng
  et~al\mbox{.}}{2018}]%
        {Zheng}
\bibfield{author}{\bibinfo{person}{Serena Zheng}, \bibinfo{person}{Noah
  Apthorpe}, \bibinfo{person}{Marshini Chetty}, {and} \bibinfo{person}{Nick
  Feamster}.} \bibinfo{year}{2018}\natexlab{}.
\newblock \showarticletitle{{User Perceptions of Smart Home IoT Privacy}}.
\newblock \bibinfo{journal}{\emph{Proc. ACM Hum.-Comput. Interact.}}
  \bibinfo{volume}{2}, \bibinfo{number}{CSCW}, Article \bibinfo{articleno}{200}
  (\bibinfo{date}{Nov.} \bibinfo{year}{2018}), \bibinfo{numpages}{20}~pages.
\newblock
\showISSN{2573-0142}
\urldef\tempurl%
\url{https://doi.org/10.1145/3274469}
\showDOI{\tempurl}


\end{thebibliography}


\end{document}